\documentclass[aps,preprint,amsmath,amssymb,amsfonts,nofootinbib,longbibliography,superscriptaddress]{revtex4-1}
\usepackage{epsfig}
\usepackage{graphicx}
\usepackage{dcolumn}
\usepackage{bm}
\usepackage{amsthm}
\usepackage{amsmath}
\usepackage{mathtools}
\usepackage{color}
\usepackage{hyperref}
\usepackage{xcolor}
\usepackage{enumitem}
\usepackage[margin=1.6cm]{geometry}
\usepackage[normalem]{ulem}

\newcommand{\de}{\mbox{d}}
\newcommand{\pa}{\partial}
\newcommand{\lf}{\left}
\newcommand{\rg}{\right}
\newcommand{\rmm}{\scriptscriptstyle\rm}

\newcommand{\be}{\begin{equation}}
\newcommand{\ee}{\end{equation}}

\numberwithin{equation}{section}
\renewcommand{\theequation}{\arabic{section}.\arabic{equation}}

\begin{document}

\preprint{CCTP-2024-16, ITCP-IPP 2024/16}

\title{Connecting Gravitational Perturbations: from Bertotti-Robinson to Extreme Reissner-Nordstr{\"o}m}

\allowdisplaybreaks

\author{Marco de Cesare}
\email{marco.decesare@na.infn.it}
\affiliation{Scuola Superiore Meridionale,
Largo San Marcellino 10, 80138 Napoli, Italy}
\affiliation{INFN, Sezione di Napoli, Complesso Univ. Monte S. Angelo, 80126 Napoli, Italy}

\author{Roberto Oliveri}
\email{roberto.oliveri@obspm.fr}
\affiliation{LUTH, Laboratoire Univers et Th{\'e}ories, Observatoire de Paris\\
CNRS, Universit{\'e} PSL, Universit{\'e} Paris Cit{\'e},
5 place Jules Janssen, 92190 Meudon, France}

\author{Achilleas P. Porfyriadis}
\email{porfyriadis@physics.uoc.gr}
\affiliation{Crete Center for Theoretical Physics, Institute of Theoretical and Computational Physics, \\ Department of Physics, University of Crete, 70013 Heraklion, Greece}

%
%
\begin{abstract}

We study spherically symmetric spacetime perturbations induced by a neutral scalar in the near-horizon region of extreme Reissner-Nordstr{\"o}m black holes. For the unperturbed black hole, the near-horizon region is given by another exact solution of the Einstein-Maxwell equations, namely the Bertotti-Robinson spacetime.
Our aim is to extend this connection beyond the background level and identify perturbations of a Bertotti-Robinson spacetime as near-horizon perturbations of an extreme Reissner-Nordstr{\"o}m black hole.
We explain that explicit identification of the perturbative solutions to the two different backgrounds can only work in appropriate gauges. 
For this reason, we first solve the two perturbation problems in the most general spherically symmetric gauges and then find the necessary gauge conditions for matching the Reissner-Nordstr{\"o}m and Bertotti-Robinson perturbative solutions in the near-horizon limit.

\end{abstract}

\maketitle

\section{Introduction}\label{Sec:Intro}

Two dimensional anti-de Sitter spacetime, $AdS_2$~, arises approximately in the near-horizon region of extreme black holes with various asymptotics. On the other hand, whenever $AdS_2$ makes such an appearance it defines exact solutions of the same Einstein equations which are asymptotically $AdS_2$ \cite{Kunduri:2007vf}. The prototypical example of this phenomenon is the pair of exact electrovacuum solutions of the Einstein-Maxwell equations in four dimensions consisting of the extreme Reissner-Nordstr{\"o}m (ERN) black hole and the Bertotti-Robinson (BR) universe \cite{Carter2009}. The situation at hand raises a natural question: is it possible to also identify perturbative solutions around the $AdS_2$ backgrounds with near-horizon approximations of perturbative solutions around the corresponding extreme black holes? 

This type of connection problem was solved in Refs.~\cite{Porfyriadis:2018yag,Porfyriadis:2018jlw} for propagating coupled gravitational and electromagnetic waves of BR and ERN. It is also easy to solve the connection problem for a linear scalar on fixed BR and ERN backgrounds. In this paper, we work out the connection of spherically symmetric gravito-electromagnetic perturbations of BR induced by a linear scalar to the corresponding reaction to such a scalar on ERN. Specifically, we consider an arbitrary spherically symmetric scalar of order $\mathcal{O}(\epsilon^{1/2})$ that solves the Klein-Gordon equation on BR or ERN in the test field approximation and the corresponding $\mathcal{O}(\epsilon)$ perturbations to the electrovacuum geometry that are induced as a reaction to it.

An important ingredient for establishing the connection of BR perturbations to ERN perturbations is the identification of appropriate gauge(s) in which such a connection is possible. Perhaps somewhat surprisingly, this aspect of the problem is particularly pronounced in the spherically symmetric case (compared e.g.~to the higher multipole waves of Refs.~\cite{Porfyriadis:2018yag,Porfyriadis:2018jlw}). The issue can be illustrated simply by the following considerations. In BR the size of the transverse $S^2$ is fixed and therefore for spherically symmetric metric perturbations their angular components are gauge-invariant. On the other hand, in ERN foliated by $S^2$'s of varying radii it is possible---in fact rather convenient---to work out the perturbative solutions in a gauge in which the angular components of the perturbation are zero. It is obvious then that such a choice of gauge for the ERN problem will make it impossible to recover the BR perturbative solutions via a near-horizon limit. Therefore, in this paper we first obtain the solutions to both perturbation problems, around BR and ERN, in the most general spherically symmetric gauges and then show how to recover the BR solutions from the near-horizon ERN solutions in appropriate gauges.

The $\mathcal{O}(\epsilon)$ gravitational perturbations we work with are a reaction to the scalar perturbation of the backgrounds that appears at the lower order $\mathcal{O}(\epsilon^{1/2})$ and we do not consider higher-order gravitational backreaction.
On the BR side, the question of backreaction of $AdS_2$ perturbations is a very interesting and subtle one. Beginning with Ref.~\cite{Maldacena:1998uz}, it was realized that backreacting perturbations in $AdS_2$ destroys the asympotically $AdS_2$ boundary conditions.
This is evident in the $\mathcal{O}(\epsilon)$ reaction perturbations
we obtain as well, which are not subleading near the $AdS_2$ boundary. In recent years it has been understood that for consistent $AdS_2$ backreaction calculations one needs to slightly break away from $AdS_2$ and study the problem in \emph{nearly}-$AdS_2$ spacetimes \cite{Almheiri:2014cka,Jensen:2016pah,Maldacena:2016upp,Engelsoy:2016xyb} (see the reviews \cite{Mertens:2022irh,Sarosi:2017ykf}).\footnote{Most of this work has been in the context of two-dimensional models of gravity, such as the Jackiw-Teitelboim theory \cite{Teitelboim:1983ux,Jackiw:1984je}. Work that emphasizes the relation to four-dimensional gravity includes Refs.~\cite{Almheiri:2016fws,Nayak:2018qej,Moitra:2018jqs}, and efforts to generalize to the context of rotating black holes include Refs.~\cite{Castro:2018ffi,Moitra:2019bub,Castro:2021csm}.}
In our context, whenever we think of the BR solutions as near-horizon approximations of ERN ones, consistent backreaction is ensured by the connection that is maintained between near-horizon and far asymptotically flat region of Reissner-Nordstr{\"o}m. The situation is rather subtle as part of the perturbative BR solution is there such that, upon backreaction, it can build the background ERN, while the remaining part is indeed the near-horizon portion of the perturbative ERN solution.
 This phenomenon of $AdS_2$ backreaction with boundary condition change, dubbed \emph{anabasis}, has been elucidated in Ref.~\cite{Hadar:2020kry}.

The paper is organized as follows. In Section~\ref{Sec:Setup} we review the dynamics, including perturbations, of the Einstein-Maxwell-scalar system in spherical symmetry. In Section~\ref{Sec:pertERN} we solve the perturbative equations in ERN in the most general spherically symmetric gauge. Similarly, the dynamical equations for perturbations in BR are solved in Section~\ref{Sec:pertBR}. Our main results are contained in Section~\ref{Sec:Matching}, where we match gravitational, electromagnetic, and matter perturbations in the two spacetimes in the near-horizon limit. Lastly, we review our results and discuss future developments in Section~\ref{Sec:Conclusion}. A technical appendix is also included, where we analyze the dynamics of the scalar field in ERN with the technique of matched asymptotic expansions.

\section{Einstein-Maxwell-scalar system in spherical symmetry}\label{Sec:Setup}
The field equations read
\begin{equation}\label{Eq:EMKGfieldeq}
G_{\mu\nu}=8\pi G \lf(T^{\rmm EM}_{\mu\nu}+T^{\phi}_{\mu\nu}  \rg)~,
\end{equation}
where
\begin{align}
T^{\phi}_{\mu\nu}&=\pa_\mu\phi\pa_\nu\phi-\frac{1}{2}(g^{\alpha\beta}\pa_\alpha\phi\pa_\beta\phi)g_{\mu\nu}~,\\
T^{\rmm EM}_{\mu\nu}&=\frac{1}{4\pi}\lf(F_{\mu}^{\;\lambda}F_{\nu\lambda}-\frac{1}{4}g_{\mu\nu}F_{\alpha\beta}F^{\alpha\beta}\rg)~,
\end{align}
with $F_{\mu\nu}=\pa_\mu A_{\nu}-\pa_\nu A_{\mu}$~. From the definition of the electromagnetic field strength follow the Bianchi identities $\nabla_{[\lambda} F_{\mu\nu]}=0$~. Since the scalar field is real, there is no direct coupling with the Maxwell field. Therefore, their stress-energy tensors are conserved separately, $\nabla^\mu T^{\phi}_{\mu\nu}=0$~, $\nabla^\mu T^{\rmm EM}_{\mu\nu}=0$~. These equations respectively imply the free wave equation $\Box\phi=0$ and the source-free Maxwell equations $\nabla^{\nu} F_{\mu\nu}=0$~. 

In the following, we adopt the framework of Refs.~\cite{Gerlach:1980tx,Gerlach:1979rw,Gundlach:1999bt} for the 2+2 decomposition of the field equations in spherical symmetry and the derivation of dynamical equations for perturbations. The latter are fully covariant on a two-dimensional Lorentzian manifold.

\subsection{General 2+2 decomposition of the field equations}\label{Sec:2+2decomposition}

We assume exact spherical symmetry for all dynamical fields (i.e., both matter and gravity). Assuming that the spacetime $(M^4,\,g_{\mu\nu})$ is spherically symmetric, the manifold $M^4$ can be factorized globally as $M^4=M^2\times S^2$~, where $M^2$ is a two-dimensional manifold and $S^2$ is the unit two-sphere. Henceforth, indices in $M^2$ are raised and lowered with $g_{AB}$~, while lower case indices in $S^2$ are raised and lowered with $\gamma_{ab}$~.
We write the metric in a general coordinate system that makes the symmetry manifest
\be\label{Eq:metric_general}
\de s^2=g_{\mu\nu}(x^{\lambda})\de x^\mu\de x^\nu= g_{AB}(x^C)\de x^A\de x^B+ r^2(x^C)\de\Omega^2~,
\ee
where capital Latin indices denote coordinates $\{x^0,x^1\}$ in $M^2$ and lower case indices represent coordinates on $S^2$~,
 $\de\Omega^2=\gamma_{ab}(x^c)\de x^a\de x^b$ is the round metric on $S^2$~, and $r$ is the areal radius.  We also define the Levi-Civita connection associated with $g_{AB}$ on $M^2$ and denote it with a vertical dash, $g_{AB|C}=0$~.
In analogy with Eq.~\eqref{Eq:metric_general}, the stress-energy tensor of a generic matter field can also be decomposed as
\be\label{Eq:stressenergy_general}
T_{\mu\nu}(x^{\lambda})\de x^{\mu}\de x^{\nu}= T_{AB}(x^C)\de x^{A}\de x^{B}+p_\perp(x^C)r^2(x^C)\de\Omega^2~,
\ee
where $p_\perp\equiv\frac{1}{2}T_d^{\;d}$ is the anisotropic pressure. Components $T_{Ab}$ must be zero as a consequence of spherical symmetry.

A common gauge choice is $x^1=r$~. Although this may be a convenient gauge choice in some case, it is not valid in the Bertotti-Robinson spacetime, where the areal radius is constant. For this reason, in the remainder of this section we will work in a general gauge.

In spherical symmetry, the scalar field is $\phi=\phi(x^C)$ and, assuming there is no magnetic charge, the Maxwell field only has a non-zero radial electric component $F_{AB}=-{\cal F}(x^C)\, \epsilon_{AB}$~, where $\epsilon_{AB}$ is the natural volume element on $M^2$~. This ansatz for the electromagnetic field automatically satisfies $\nabla_{[\lambda} F_{\mu\nu]}=0$~. The remaining Maxwell equations $\nabla^{\nu} F_{\mu\nu}=0$ boil down to $(r^2F_{AB})^{|B}=0$~, which implies the Coulombian form for the electric field
\be\label{Eq:Maxwellfieldsol}
{\cal F}(x^C)=\frac{Q}{r^2(x^C)}~,
\ee
where the electric charge $Q$ is a constant.

The non-zero components of the stress-energy tensor of the scalar field read
\be\label{Eq:Tscalar_decompose}
T^{\phi}_{AB}=\pa_A\phi\pa_B\phi-\frac{1}{2}(g^{CD}\pa_C\phi\pa_D\phi)g_{AB}~,\quad p^{\phi}_\perp=-\frac{1}{2}(g^{CD}\pa_C\phi\pa_D\phi)~.
\ee
For the Maxwell field, using the solution \eqref{Eq:Maxwellfieldsol}, we have
\be\label{Eq:Tmaxwell_decompose}
T^{\rmm EM}_{AB}=-\frac{Q^2}{8\pi r^4}g_{AB}~,\quad p^{\rmm EM}_\perp=\frac{Q^2}{8\pi r^4}~.
\ee

Following \cite{Gerlach:1980tx,Gerlach:1979rw,Gundlach:1999bt}, the non-zero components of the Einstein field equations \eqref{Eq:EMKGfieldeq} can be decomposed as
\begin{subequations}\label{Eq:fieldeq_sphericalsym}
\begin{align}
G_{AB}=-2(v_{A|B}+v_Av_B)+g_{AB}\lf( 2v_C^{\phantom{C}|C}+3v_Cv^C-\frac{1}{r^2}\rg)&=8\pi G \lf(T^{\phi}_{AB} -\frac{Q^2}{8\pi r^4}g_{AB}\rg)~,\label{Eq:fieldeq_sphericalsym_a}\\
\frac{1}{2}G_a^{\phantom{a}a}=-\frac{1}{2}R^{(2)}+v_A^{\phantom{A}|A}+v_Av^A&=8\pi G \lf(p^{\phi}_\perp+\frac{Q^2}{8\pi r^4}\rg) ~,\label{Eq:fieldeq_sphericalsym_b}
\end{align}
\end{subequations}
where $R^{(2)}$ is the Ricci scalar of $g_{AB}$ in $M^2$ and $v_A\equiv \pa_Ar/r$~. The Bianchi identities imply the continuity equation
\be\label{Eq:continuity_decomposed}
\frac{1}{r^2}\lf(r^2 T^{\phi}_{AB} \rg)^{|B}=2p^{\phi}_\perp v_A~,
\ee
which, using Eq.~\eqref{Eq:Tscalar_decompose}, can be recognized as equivalent to the wave equation $\Box\phi=0$~.
Alternatively, Eq.~\eqref{Eq:fieldeq_sphericalsym_b} can be derived from Eq.~\eqref{Eq:fieldeq_sphericalsym_a} and the dynamical equation for the scalar. In fact, multiplying by $r^2$ and taking the covariant derivative of Eq.~\eqref{Eq:fieldeq_sphericalsym_a} on $M^2$~, and after substituting  Eq.~\eqref{Eq:continuity_decomposed}, one obtains
\be
\lf[-\frac{1}{2}R^{(2)}+v_C^{\phantom{C}|C}+v_Cv^C-8\pi G \lf(p^{\phi}_\perp+\frac{Q^2}{8\pi r^4}\rg)\rg]v_A=0~.
\ee
This is equivalent to Eq.~\eqref{Eq:fieldeq_sphericalsym_b} {\it if} $v_A\neq0$~.
Therefore, in a generic spherically symmetric spacetime where $v_{A}\neq0$~, Eq.~\eqref{Eq:fieldeq_sphericalsym_b} can be regarded as a consequence of \eqref{Eq:fieldeq_sphericalsym_a} supplemented by the equation of motion for $\phi$~.

However, there is one important exception to the argument given above: if $v_A=0$ identically, that is the areal radius is constant $r=r_0$~, then Eqs.~\eqref{Eq:fieldeq_sphericalsym_a},~\eqref{Eq:fieldeq_sphericalsym_b} are independent of each other and boil down, respectively, to
\be\label{Eq:FieldEqsBR}
\frac{GQ^2-r_0^2}{r_0^4}g_{AB}=8\pi G\,T_{AB}^{\phi} ~,\quad -\frac{1}{2}R^{(2)}=8\pi G p^{\phi}_\perp+ \frac{G Q^2}{r_0^4}~.
\ee
The first equation in \eqref{Eq:FieldEqsBR} implies that $r_0^2=G Q^2$ and $\phi=constant$~. This solution is known as the Bertotti-Robinson spacetime, with $M^2\cong AdS_2$~, whose curvature radius is the reciprocal of the $S^2$ radius. We will see in the following Section~\ref{Sec:Perturbations_General} that, at the perturbative level as well, whenever Bertotti-Robinson is assumed as a background, we need to take into account Eq.~\eqref{Eq:fieldeq_sphericalsym_b} separately.

\subsection{Perturbed field equations}\label{Sec:Perturbations_General}
Our perturbative scheme is as follows. We consider a scalar field propagating on an electrovacuum background and compute the reaction of the geometry to it. That is to say, we assume that gravitational perturbations and the stress-energy tensor of the scalar field are both of the order of some small parameter $\epsilon$~.

In this section we denote the perturbed metric as 
\be
\overline{g}_{\mu\nu}(x^C)=g_{\mu\nu}(x^C)+h_{\mu\nu}(x^C)~,
\ee
where $g_{\mu\nu}$ is the background and $h_{\mu\nu}$ the perturbation.
We assume spherically symmetric perturbations, that is the only non-vanishing components of the metric perturbations are $h_{AB}(x^C)$ and $h_{ab}(x^C)\equiv r^2(x^C) K(x^C) \gamma_{ab}$~.
More explicitly
\be
\overline{g}_{AB}=g_{AB}+h_{AB}~,\quad \overline{r^2}=r^2(1+K)~.
\ee
For generality, we also include a perturbation of the electric charge (even though this cannot be sourced dynamically by the real scalar field)
\be
\overline{Q}=Q+\delta Q~.
\ee

In the following we linearize Eqs.~\eqref{Eq:fieldeq_sphericalsym_a}, \eqref{Eq:fieldeq_sphericalsym_b} with $h_{\mu\nu}~,\,T^{\phi}_{\mu\nu}={\cal O}(\epsilon)$~. 
Notice that in the aforementioned non-linear equations we make the necessary notational shift: we replace all quantities with their barred counterparts ($\overline{g}_{\mu\nu},~\overline{v}_A,~\overline{r}$, etc) and proceed expanding. As a result, from now on unbarred quantities ($g_{\mu\nu},~v_A,~r$, etc) will refer to the background.

\subsubsection{Perturbations on a general spherically symmetric background}
Linearizing Eqs.~\eqref{Eq:fieldeq_sphericalsym_a}, \eqref{Eq:fieldeq_sphericalsym_b} around a general spherically symmetric electrovacuum background, we obtain
\begin{subequations}
\be\label{Eq:fieldeq_sphericalsym_elimEM_perturbed}
\begin{split}
&\lf(h_{AC|B}+h_{BC|A}-h_{AB|C}-2g_{AB}h_{CD}^{\phantom{CD}|D}+g_{AB}h^{D}_{\phantom{D}D|C}\rg)v^C+h_{AB}{\cal V}\\
&\quad -g_{AB}\lf(3h_{CD}v^Cv^D+2h_{CD}v^{C|D}\rg)-\lf(K_{|AB}+K_{|A}v_B+K_{|B}v_A\rg)\\
&\quad +g_{AB}\lf(3K_{|C}v^C+K_{|C}^{\phantom{|C}C}+\frac{r^2-2GQ^2}{r^4}K+\frac{2GQ\delta Q}{r^4}\rg)=8\pi G \,T^{\phi}_{AB}~, 
\end{split}
\ee
\be\label{Eq:fieldeq_angular_perturbed}
\begin{split}
-h^{AB}_{\phantom{AB}|AB}+h^{A\phantom{A|}B}_{\phantom{A}A|\phantom{B}B}-2h^{AB}_{\phantom{AB}|A}v_B+h^A_{\phantom{A}A|B}v^B-2h^{AB}v_{A|B}-2h^{AB}v_Av_B\\
\qquad +\frac{R^{(2)}}{2}h^A_{\phantom{A}A}+ K_{|A}^{\phantom{A}A}+2K_{|A}v^A+\frac{4GQ^2}{r^4}K=16\pi G\, p_{\phi}+\frac{4GQ\delta Q}{r^4}~.  
\end{split}
\ee
\end{subequations}
In the equations above, $R^{(2)}$~, $v_A$~, and covariant derivatives are defined with respect to background quantities, and we defined ${\cal V}\equiv-\frac{1}{r^2}+2v_{C}^{\phantom{C}|C}+3v^Cv_C+\frac{GQ^2}{r^4}$~. The perturbative equations above are fully covariant on $M^2$~.

The scalar field propagates as a test field on the background geometry, and as such it is subject to $\Box \phi=0$~, and Eqs.~\eqref{Eq:fieldeq_sphericalsym_elimEM_perturbed}, \eqref{Eq:fieldeq_angular_perturbed} allow us to compute the reaction of the geometry it.

\subsubsection{Perturbations on a Bertotti-Robinson background}
In a Bertotti-Robinson geometry, we have $r^2=r_0^2=G Q^2$ identically and therefore $v_A=0$~. This also implies that ${\cal V}=0$~. Thus, equations \eqref{Eq:fieldeq_sphericalsym_elimEM_perturbed}, \eqref{Eq:fieldeq_angular_perturbed} simplify considerably:
\begin{subequations}
\begin{align}
&g_{AB}\lf(K_{|C}^{\phantom{|C}C}-\frac{1}{r_0^2}K+\frac{2\delta Q}{r_0^2Q}\rg)-K_{|AB}=8\pi G T^{\phi}_{AB}~,
 \label{Eq:PertBRcovariant1}\\
&-h^{AB}_{\phantom{AB}|AB}+h^{A\phantom{A|}B}_{\phantom{A}A|\phantom{B}B}-\frac{1}{r_0^2}h^A_{\phantom{A}A}+
K_{|A}^{\phantom{A}A}+\frac{4}{r_0^2}K=16\pi G p_{\phi}+\frac{4\delta Q}{r^2_0Q}~. \label{Eq:PertBRcovariant2}
\end{align}
\end{subequations}
We note that the $M^2$ metric perturbation, $h_{AB}$~, does not appear in Eq.~\eqref{Eq:PertBRcovariant1}. Hence, in this special case the angular component, Eq.~\eqref{Eq:PertBRcovariant2}, is an independent equation and must also be taken into account.
A good strategy to solve these equations is as follows. Begin by taking the trace of Eq.~\eqref{Eq:PertBRcovariant1}, to obtain
\be\label{Eq:kleingordonKpert}
K_{|C}^{\phantom{|C}C}-\frac{2}{r_0^2}K+\frac{4\delta Q}{r_0^2Q}=0~, 
\ee
which is the Klein-Gordon equation for a scalar field with mass $\mu^2=2/r_0^2$~, including a constant source $\propto\delta Q/Q$~.
Then substitute Eq.~\eqref{Eq:kleingordonKpert} into Eq.~\eqref{Eq:PertBRcovariant2} 
\be\label{Eq:leftovereqBR}
h^{AB}_{\phantom{AB}|AB}-h^{A\phantom{A|}B}_{\phantom{A}A|\phantom{B}B}+\frac{1}{r_0^2}h^A_{\phantom{A}A}=
\frac{6}{r_0^2}K-16\pi G p_{\phi}-\frac{8\delta Q}{r^2_0Q}~. 
\ee
In this equation $K$ plays the role of a source for $h_{AB}$~. Once Eq.~\eqref{Eq:PertBRcovariant1} has been solved, Eq.~\eqref{Eq:leftovereqBR} can be used to determine the remaining independent component of the metric perturbation. This is the strategy that will be adopted in the following sections.

\subsection{Gauge transformations}
The transformation of the perturbed metric under an infinitesimal diffeomorphism generated by a vector field $\xi^{\mu}$ is $\overline{g}_{\mu\nu}\rightarrow \overline{g}_{\mu\nu}-\pounds_\xi \overline{g}_{\mu\nu}$~. Expanding the metric $\overline{g}_{\mu\nu}$ as background plus perturbations, we get the following transformation rule for the latter
\be\label{Eq:DiffeoDef}
\Delta h_{\mu\nu}=-(\pounds_\xi g)_{\mu\nu}=-\xi_{\mu;\nu}-\xi_{\nu;\mu}~.
\ee
(Here and in the following $\Delta$ represents the effect of an infinitesimal gauge transformation on a given field.)
In particular, we are interested in infinitesimal diffeomorphisms that preserve spherical symmetry (i.e.,~the structure \eqref{Eq:metric_general} for the metric). The corresponding generators have zero angular components, and thus can be expanded as
\be\label{Eq:DiffeoGenDef}
\xi_\mu\de x^\mu= \xi_A \de x^A~.
\ee
Consider diffeomorphisms with $\xi_A=\mathcal{O}(\epsilon)$~.
Performing the spherical decomposition of Eq.~\eqref{Eq:DiffeoDef} and using Eq.~\eqref{Eq:DiffeoGenDef} we obtain
\begin{subequations}\label{Eq:GaugeTransformationKhpert}
\begin{align}
\Delta h_{AB}&= -\xi_{A|B}-\xi_{B|A}~\\
\Delta K&=-2 v^A\xi_A~.\label{Eq:GaugeTransformationK}
\end{align}
\end{subequations}
Similarly, the components of the stress-energy tensor \eqref{Eq:stressenergy_general} transform as
\begin{subequations}
\begin{align}
\Delta T_{AB}&=-T_{AB|C}\xi^C-T_{AC}\xi^C_{\phantom{C}|B}-T_{BC}\xi^C_{\phantom{C}|A}~,\\
\Delta p_{\perp}&= -\frac{1}{r^2}\lf(r^2 p_{\perp}\rg)_{|A}\xi^A~.
\end{align}
\end{subequations}
The transformation properties of the electromagnetic stress-energy tensor can be computed using the above. The scalar field transforms as $\Delta\phi = - \phi_{|A}\xi^A$~.
Moreover, since we are assuming that the scalar field stress-energy tensor does not contribute at the background level and is therefore $\mathcal{O}(\epsilon)$~, we have $\Delta T^{\phi}_{AB}=\mathcal{O}(\epsilon^2)$~. The Maxwell field transforms as $\Delta {\cal F}=- ({\cal F} \xi^A)_{|A}$ under infinitesimal diffeomorphisms, and it is gauge-invariant under the electromagnetic $U(1)$ gauge symmetry.

Note that Eq.~\eqref{Eq:GaugeTransformationK} implies that, if so desired, the perturbation $K$ can be gauged away for all spherically symmetric spacetimes with $v_A\neq0$~, that is to say in all but the Bertotti-Robinson case. This observation will be relevant in later sections.

\subsection{Misner-Sharp mass}
The Misner-Sharp mass function is
\be\label{Eq:MisnerSharpDefinition}
m_{\rm MS}(x^C)\equiv\frac{r}{2G}\lf(1-g^{AB}\pa_A r\pa_B r \rg)=\frac{r}{2G}\lf(1-r^2 v_A v^A  \rg)~.
\ee
The perturbation of $m_{\rm MS}$ in a generic gauge reads
\be\label{Eq:MisnerSharpPerturbation}
\delta m_{\rm MS}=\frac{r}{4G}\lf[ K(1-3r^2v^Av_A)+2r^2\lf(h_{AB}v^Av^B-v^AK_{|A}\rg)\rg]~.
\ee
Using the transformation properties \eqref{Eq:GaugeTransformationKhpert}, under infinitesimal gauge transformations the Misner-Sharp mass transforms as
\be
\Delta m_{\rm MS}=-\frac{r}{2G}\lf[v^C\xi_C(1-3r^2v^Av_A)-2r^2\lf(v^Av^B_{\phantom{B}|A}\xi_{B}\rg) \rg]~.
\ee
Thus, in the general case the perturbation $\delta m_{\rm MS}$ is not gauge invariant under infinitesimal diffeomorphisms, except on a Bertotti-Robinson background (where $v_A=0$ identically).

\section{Perturbations of extremal Reissner-Nordstr{\"o}m}\label{Sec:pertERN}

In this section we study the dynamics of the perturbations of an extremal Reissner-Nordstr{\"o}m (ERN) spacetime. The background metric, in coordinates such that the event horizon is at $\rho=0$~, may be written as
\begin{equation}\label{Eq:Background_ERN}
    \de s^2=(G M_0)^2\lf[- \lf(\frac{\rho}{1+\rho}\rg)^2\de \tau^2 +\lf(\frac{1+\rho}{\rho}\rg)^2\de \rho^2+(1+\rho)^2\de\Omega^2 \rg]~,
\end{equation}
where $M_0$ is the ADM mass, which is related to the extremal electric charge value by $Q^2=GM_0^2$~. The coordinates $(\tau,\rho)$ are dimensionless. Comparing with the notation used in the previous section for the most general spherically symmetric spacetime, here we have made the choice $x^0=\tau$~, $x^1=\rho$~, and $r=G M_0(1+\rho)$~.
 The background solution for the Maxwell field in these coordinates reads ${\cal F}(\rho)=Q/r_0^2(1+\rho)^2$~, where we have defined $r_0^2=GQ^2$~, such that the near-horizon limit of the line element \eqref{Eq:Background_ERN}  yields a Bertotti-Robinson spacetime with curvature radius $r_0$~. In the following we will assume $Q>0$ for definiteness. The background solution for the scalar field is $\phi=constant$~, corresponding to zero stress-energy.

Given the form of the background metric \eqref{Eq:Background_ERN}, it is convenient to rescale the perturbations as $h_{AB}=r_0^2 H_{AB}$~.
We work in a fully general gauge for the perturbations, i.e.~we do not make any specific assumptions for the components of $H_{AB}$~.

The perturbed Maxwell field is obtained from Eq.~\eqref{Eq:Maxwellfieldsol}
\be\label{Eq:PertMaxwellERN}
\overline{{\cal F}}=\frac{\overline{Q}}{\overline{r}^2}=\frac{Q+\delta Q^{\rmm ERN}}{r_0^2(1+\rho)^2(1+K)}~.
\ee
The scalar field obeys the wave equation in the ERN background
\be\label{Eq:ScalarWaveEquationERN}
-(1+\rho)^4\pa_{\tau}^2\phi+\rho^4\pa_{\rho}^2\phi+2\rho^3\pa_{\rho}\phi=0~.
\ee
The perturbed Einstein equations \eqref{Eq:fieldeq_sphericalsym_elimEM_perturbed} read, in components:
\be\label{Eq:perteqTauTauERN}
\begin{split}
&(1+\rho)^2 \rho^2 \pa_\rho^2K+\rho(1+\rho)(1+3\rho)\pa_\rho K+\lf(\rho(2+\rho)-1 \rg)K=-2\frac{\delta Q^{\rmm ERN}}{Q}\\
&-4 \pi  G \frac{(1+\rho)^2}{\rho^2}\Big( \rho^4  (\pa_\rho\phi)^2+(1+\rho)^4 (\pa_\tau\phi)^2\Big)+(1+\rho)^2\pa_\rho\lf(\frac{\rho^4}{(1+\rho)^3}H_{\rho\rho} \rg)~,
\end{split}
\ee
\be\label{Eq:perteqTauRhoERN}
\rho(1+\rho)\pa_{\tau}\pa_{\rho}K-(1-\rho)\pa_{\tau}K= - 8 \pi  G\rho (1+\rho) \pa_\rho\phi \pa_\tau\phi+\frac{\rho^3}{(1+\rho)^2}\pa_{\tau}H_{\rho\rho}~,
\ee
\begin{align}\label{Eq:perteqRhoRhoERN}
&(1+\rho)^6 \pa^2_{\tau}K-\rho^3(1+\rho)^2\pa_{\rho}K+\rho^2(1-2\rho-\rho^2)K-2\rho^2(1+\rho)^3\pa_\tau H_{\tau\rho}-2\rho^2\frac{\delta Q^{\rmm ERN}}{Q}=\\
&-4 \pi  G(1+\rho)^2 \Big((1+\rho)^4 (\pa_\tau\phi)^2+\rho^4 (\pa_\rho\phi)^2\Big)- \rho^4 (1+\rho) \pa_{\rho}\lf( \frac{(1+\rho)^2} {\rho^2}H_{\tau\tau} \rg)-\frac{\rho^5(2+\rho)}{(1+\rho)^2}H_{\rho\rho}~.\nonumber
\end{align}
The angular component of the perturbed field equations \eqref{Eq:fieldeq_angular_perturbed} is
\be\label{Eq:ERNpertAngular}
\begin{split}
2\rho(1+\rho)^2\pa_\rho\lf(\rho \pa_\tau H_{\tau\rho}\rg)-(1+\rho)^4\pa_\tau^2K+\rho^4\pa_\rho^2K+2\rho^3\pa_\rho K+\frac{4\rho^2}{(1+\rho)^2}K=\\
\frac{4\rho^2}{(1+\rho)^2}\frac{\delta Q^{\rmm ERN}}{Q}+8\pi G \Big( \lf(1+\rho\rg)^4(\pa_\tau\phi)^2- \rho^4(\pa_\rho\phi)^2 \Big)+X(\tau,\rho)~,
\end{split}
\ee
with
\be
\begin{split}
X(\tau,\rho)\equiv \rho^2(1+\rho)^2\pa_{\tau}^2 H_{\rho\rho}+\frac{\rho^5}{(1+\rho)^2}\pa_{\rho}H_{\rho\rho}+\frac{2\rho^4(2+\rho)}{(1+\rho)^4}H_{\rho\rho}\\
+\rho^2(1+\rho)^2\pa_{\rho}^2 H_{\tau\tau}-\rho(1-\rho^2)\pa_\rho H_{\tau\tau}+2\rho H_{\tau\tau}~.
\end{split}
\ee
We recall that not all of the above equations are independent (see the relevant discussion in Section~\ref{Sec:2+2decomposition}).

In Eqs.~\eqref{Eq:perteqTauTauERN}, \eqref{Eq:perteqTauRhoERN}, it is convenient to change variable and define
\be\label{Eq:DefW}
W\equiv\frac{(1+\rho)^2}{\rho}K~,
\ee
in terms of which these equations read as
\be\label{Eq:perteqW1}
\begin{split}
\rho^3\pa_{\rho}^2W+\frac{\rho^2(3+\rho)}{1+\rho}\pa_\rho W=&-2\frac{\delta Q^{\rmm ERN}}{Q}+(1+\rho)^2\pa_\rho\lf(\frac{\rho^4}{(1+\rho)^3}H_{\rho\rho} \rg)\\
&-4\pi G\frac{(1+\rho)^2}{\rho^2}\Big( \rho^4  (\pa_\rho\phi)^2+(1+\rho)^4 (\pa_\tau\phi)^2\Big)~,
\end{split}
\ee
\be\label{Eq:perteqW2}
\pa_\tau\pa_\rho W=-8\pi G\frac{(1+\rho)^2}{\rho}\pa_\tau\phi\pa_\rho\phi+\frac{\rho}{1+\rho}\pa_\tau H_{\rho\rho}~.
\ee
The general solution of Eq.~\eqref{Eq:perteqW1}
is
\be\label{Eq:WsolERN}
\begin{split}
W(\tau,\rho)=&f(\tau)-\frac{(1+\rho)^2}{\rho^2}\frac{\delta Q^{\rmm ERN}}{Q}+8\pi G\, q(\tau)\int_{\delta}^{\rho} \de \rho^{\prime}\, \frac{(1+\rho^{\prime})^2}{\rho^{\prime\, 3}} +\int \de\rho\; \frac{\rho\, H_{\rho\rho}}{1+\rho}\\
&- 4\pi G\int_{\delta}^{\rho}\de \rho^{\prime}\;\frac{(1+\rho^{\prime})^2}{\rho^{\prime\,3}} \int_{\delta}^{\rho^{\prime}}\de \rho^{\prime\prime}\; \frac{1}{\rho^{\prime\prime\,2}}\Big((1+\rho^{\prime\prime})^4(\pa_\tau\phi)^2+\rho^{\prime\prime\,4}(\pa_{\rho^{\prime\prime}}\phi)^2 \Big) ~,
\end{split}
\ee
where $f(\tau)$ is a free function and $\delta>0$ an arbitrary constant. We note that the choice of antiderivatives in the indefinite integral above can be reabsorbed into $f(\tau)$~.
On the other hand, $q(\tau)$ is determined up to a constant by the dynamics of the scalar field. Indeed, substituting Eq.~\eqref{Eq:WsolERN} into \eqref{Eq:perteqW2} gives
\be\label{Eq:MomentumFlux}
\frac{\de q}{\de\tau}= -\rho^2 \pa_\tau\phi \pa_{\rho}\phi+\frac{\pa}{\pa\tau}\int_{\delta}^{\rho}\de \rho^{\prime}\; \frac{1}{2\rho^{\prime\,2}}\Big((1+\rho^{\prime})^4(\pa_\tau\phi)^2+\rho^{\prime\,4}(\pa_{\rho^{\prime}}\phi)^2 \Big)~.
\ee
Given Eq.~\eqref{Eq:ScalarWaveEquationERN}, it is straightforward to show that the right hand side of the above is purely a function of $\tau$~.
In fact, equation~\eqref{Eq:MomentumFlux} can also be rewritten more geometrically as
\be
\frac{\de q}{\de\tau}+r^2(\rho) T^{\rho}_{\;\tau}=\frac{\pa}{\pa\tau}\int_{\delta}^{\rho}\de x\; r^2(x) (-T^{\tau}_{\;\tau})~,
\ee
which makes apparent its interpretation as a continuity equation, with $-\de q(\tau)/\de\tau$ being the energy-momentum flux through the surface $r=\delta$~.

It remains to solve Eqs.~\eqref{Eq:perteqRhoRhoERN} and \eqref{Eq:ERNpertAngular}. In order to prepare the ground for comparing with the perturbative solution in BR, it is convenient to first eliminate second order derivatives of $K$ from \eqref{Eq:ERNpertAngular} and then solve for $H_{\tau\rho}$~. Specifically, using Eqs.~\eqref{Eq:perteqTauTauERN}, \eqref{Eq:perteqRhoRhoERN} in Eq.~\eqref{Eq:ERNpertAngular}, we obtain 
\be\label{Eq:ERNpertAngular_simpl}
\begin{split}
\rho(1+\rho)^4\pa_{\rho}\pa_{\tau}H_{\tau\rho}+(1+\rho)^3\pa_{\tau}H_{\tau\rho}+(3-2\rho-\rho^2)\rho K -\rho^3(1+\rho)\pa_\rho K + Y(\tau,\rho)=\\
4\pi G\frac{(1+\rho)^2}{\rho}\Big((1+\rho)^4\lf(\pa_{\tau}\phi\rg)^2-\rho^4\lf(\pa_{\rho}\phi\rg)^2 \Big)
+4\rho\frac{\delta Q^{\rmm ERN}}{Q}~,
\end{split}
\ee
with
\be
\begin{split}
 Y(\tau,\rho)\equiv-\frac{1}{2}\lf[\rho(1+\rho)^4\pa_{\rho}^2H_{\tau\tau}-(1+\rho)^3\pa_{\rho}H_{\tau\tau}+\rho(1+\rho)^4\pa_{\tau}^2H_{\rho\rho}+\frac{\rho^4}{1+\rho}\pa_\rho H_{\rho\rho}\rg.\\
\lf.+\frac{2\rho^3(2-2\rho-\rho^2)}{(1+\rho)^2}H_{\rho\rho}+4(1+\rho)^2H_{\tau\tau}\rg]~.
\end{split}
\ee
The solution to the above reads
\be\label{Eq:ERNsol_tr_Simp}
\begin{split}
&H_{\tau\rho}(\tau,\rho)=s(\rho)-\frac{1}{\rho(1+\rho)^2}\frac{\delta Q^{\rmm ERN}}{Q}\tau\\
&-\frac{\rho^2}{(1+\rho)^5}\int_0^{\tau}\de\tau^{\prime}\, f(\tau^{\prime})+\frac{1+\rho}{\rho}\int_0^{\tau}\de\tau^{\prime}\, j(\tau^{\prime})+\frac{1+\rho}{\rho}\int_0^{\tau}\de\tau^{\prime} \int\de\rho\,  Z(\tau^{\prime},\rho) \\
&+8\pi G \frac{(1+\rho)}{\rho}\lf[\int_{\delta}^{\rho}\de\rho^{\prime}\, \Bigg(\frac{\rho^{\prime}}{(1+\rho^{\prime})^4}-\frac{3\rho^{\prime\,2}(1-\rho^{\prime})}{(1+\rho^{\prime})^7}\int_{\delta}^{\rho^{\prime}}\de\rho^{\prime\prime}\,\frac{(1+\rho^{\prime\prime})^2}{\rho^{\prime\prime\,3}}  \Bigg)\rg] \int_0^{\tau}\de\tau^{\prime}\,  q(\tau^{\prime})\\
&-\frac{4\pi G (1+\rho)}{\rho}\int_{0}^{\tau}\de\tau^{\prime}\int_{\delta}^{\rho}\de\rho^{\prime}\lf\{\frac{\rho^{\prime\,4}(\pa_{\rho^{\prime}}\phi)^2-(1+\rho^{\prime})^4(\pa_{\tau^{\prime}}\phi)^2}{\rho^{\prime}(1+\rho^{\prime})^3}  \rg.\\
&\hspace{13em}+\frac{\rho^{\prime}}{(1+\rho^{\prime})^4}\int_{\delta}^{\rho^{\prime}}\de\rho^{\prime\prime}\frac{(1+\rho^{\prime\prime})^4(\pa_{\tau^{\prime}}\phi)^2+\rho^{\prime\prime\;4}(\pa_{\rho^{\prime\prime}}\phi)^2}{\rho^{\prime\prime\; 2}}\\
&\qquad\qquad\lf. -\frac{3(1-\rho^{\prime})\rho^{\prime\,2}}{(1+\rho^{\prime})^7}\int_{\delta}^{\rho^{\prime}}\de\rho^{\prime\prime}\frac{(1+\rho^{\prime\prime})^2}{\rho^{\prime\prime\,3}}\int_{\delta}^{\rho^{\prime\prime}}\de\rho^{\prime\prime\prime}\frac{(1+\rho^{\prime\prime\prime})^4(\pa_{\tau^{\prime}}\phi)^2+ \rho^{\prime\prime\prime\,4}(\pa_{\rho^{\prime\prime\prime}}\phi)^2}{\rho^{\prime\prime\prime\,2}} \rg\}~,
\end{split}
\ee
with
\be\label{Eq:DefXtripletilde}
\begin{split}
 Z(\tau,\rho)\equiv\frac{2(1-\rho)\rho^3}{(1+\rho)^7}H_{\rho\rho}+\frac{\rho^4 \pa_{\rho}H_{\rho\rho}}{2(1+\rho)^6}+\frac{\rho \pa_{\tau}^2 H_{\rho\rho}}{2(1+\rho)}+\frac{\rho \pa_{\rho}^2 H_{\tau\tau}}{2(1+\rho)} \\
 -\frac{\pa_{\rho}H_{\tau\tau} }{2(1+\rho)^2}+\frac{2H_{\tau\tau}}{(1+\rho)^3}-\frac{3(1-\rho)\rho^2}{(1+\rho)^7}\int\de\rho\, \frac{\rho H_{\rho\rho}}{1+\rho}~.
\end{split}
\ee
Here $s(\rho)$ is a free function, while $j(\tau)$ is fixed by substituting the solution \eqref{Eq:ERNsol_tr_Simp} into the remaining equation \eqref{Eq:perteqRhoRhoERN}:

\be
\begin{split}
2j(\tau)=&\; \frac{\de^2 f(\tau)}{\de \tau^2}+\frac{S(\tau,\rho)}{(1+\rho)^6}-2\int\de\rho\,  Z(\tau,\rho)+\frac{4 \pi  G}{(1+\rho)^2}\Big(U(\tau,\rho)+\rho\, V(\tau,\rho)-2q(\tau)\Big) \\
&+4\pi G  \lf(\int_\delta^\rho \de\rho^{\prime}\, \frac{(1+\rho^{\prime})^2}{\rho^{\prime\,3}}\rg)  \lf(\pa_{\tau}^2U(\tau,\rho)-2  \frac{\pa}{\pa\tau}\lf(\rho^2\pa_{\tau}\phi\pa_{\rho}\phi\rg) \rg)\\
&+ \frac{ 8\pi G\, \rho^3 }{(1+\rho)^6}\int _{\delta}^\rho \de \rho^{\prime}\,\frac{(1+\rho^{\prime})^2}{\rho^{\prime\,3}} U(\tau,\rho^{\prime})- 4\pi G \int _{\delta}^\rho \de \rho^{\prime}\,\frac{(1+\rho^{\prime})^2}{\rho^{\prime\,3}} \pa_{\tau}^2 U(\tau,\rho^{\prime}) \\
     &+8\pi G\, \int_{\delta}^\rho\, \de\rho^{\prime} \lf[\frac{\rho^{\prime}\, U(\tau,\rho^{\prime})}{(1+\rho^{\prime})^4}-\frac{2q(\tau)}{(1+\rho^{\prime})^3}+\frac{\rho^{\prime\,4}(\pa_{\rho^{\prime}}\phi)^2-(1+\rho^{\prime})^4(\pa_{\tau}\phi)^2}{\rho^{\prime}(1+\rho^{\prime})^3}\rg.\\
     &\lf.-\frac{3\rho^{\prime\,2}(1-\rho^{\prime})}{(1+\rho^{\prime})^7}\int_{\delta}^{\rho^{\prime}} \de \rho^{\prime\prime}\,\frac{(1+\rho^{\prime\prime})^2}{\rho^{\prime\prime\,3}} U(\tau,\rho^{\prime\prime})\rg]~,
   \end{split}
   \ee
      where we defined
   \be
  V(\tau,\rho)\equiv\frac{1}{\rho^2}\Big((1+\rho)^4(\pa_\tau\phi)^2+\rho^4(\pa_{\rho}\phi)^2 \Big)~,\qquad  U(\tau,\rho)\equiv\int_{\delta}^\rho \de\rho^{\prime}\,V(\tau,\rho^{\prime})~,
   \ee
   and
\be\label{Eq:XtildeDef}
S(\tau,\rho)\equiv\rho^4 H_{\rho\rho}-2(1+\rho)^4H_{\tau\tau}+\rho(1+\rho)^5\pa_{\rho}H_{\tau\tau}-2\rho^3\int\de\rho\;\frac{\rho\,H_{\rho\rho}}{1+\rho}+(1+\rho)^6\int\de\rho\;\frac{\rho\,\pa_{\tau}^2 H_{\rho\rho}}{1+\rho} ~.
\ee

Note that in the electrovacuum case ($\phi=constant$) we have $q(\tau)= k$~, where $k$ is a constant.
As a result, the most general spherically symmetric electrovacuum perturbations of ERN are a two-parameter family of solutions parameterized here by $k$ and $\delta Q^{\rmm ERN}$~, which simply span the phase space of linear shifts in the ADM mass and electric charge of the black hole with respect to the background.

Lastly, the perturbed Misner-Sharp mass, computed in the general case using Eqs.~\eqref{Eq:MisnerSharpDefinition}, \eqref{Eq:MisnerSharpPerturbation}, reads
\be\label{Eq:MisnerSharpERN}
m_{\rm MS}=\frac{r_0}{2G(1+\rho)}\lf(1+2\rho+\frac{\rho^4}{(1+\rho)^2}H_{\rho\rho}+\lf(\frac{1}{2}+\rho-\rho^2\rg)K-\rho^2(1+\rho)\pa_\rho K \rg)~.
\ee

\section{Perturbations of Bertotti-Robinson spacetime}\label{Sec:pertBR}

In this section we consider a Bertotti-Robinson (BR) background, that is $M^4\cong AdS_2\times S^2$~, where both the $AdS_2$ and $S^2$ factors have the same curvature radius $r_0$ and the metric reads
\begin{equation}\label{Eq:Background_BR}
    \de s^2=r_0^2\lf(- \rho^2\de \tau^2 +\frac{\de \rho^2}{\rho^2}+\de\Omega^2 \rg)~.
\end{equation}
From the general solution of the Maxwell equations \eqref{Eq:Maxwellfieldsol} and the gravitational field equations~\eqref{Eq:FieldEqsBR}, it follows that the background configuration of the Maxwell field is constant, ${\cal F}=Q/r_0^2$ with $r_0^2=G Q^2$~. The background scalar field $\phi$ is also constant and therefore does not contribute to the background stress-energy tensor.
For consistency with the previous section, also in this case we assume $Q>0$ .
As before, we rescale the perturbations as $h_{AB}=r_0^2 H_{AB}$ and work in a general gauge.
Recall that on a Bertotti-Robinson background $K$ is gauge-invariant, $\Delta K=0$~, as a consequence of Eq.~\eqref{Eq:GaugeTransformationK}.

The perturbed Maxwell field reads
\be\label{Eq:PertMaxwellBR}
\overline{{\cal F}}=\frac{\overline{Q}}{\overline{r}^2}=\frac{Q+\delta Q^{\rmm BR}}{r_0^2(1+K)}~.
\ee
The scalar field obeys the wave equation in the BR background
\be\label{Eq:ScalarWaveEquationBR}
-\pa_{\tau}^2\phi+\rho^4\pa_{\rho}^2\phi+2\rho^3\pa_{\rho}\phi=0~.
\ee

The perturbed Einstein equations \eqref{Eq:fieldeq_sphericalsym_elimEM_perturbed} read, in components:
\begin{subequations}\label{EQ:BRpertAll}
\begin{align}
&\rho^2 \pa_\rho^2K+\rho\pa_\rho K-K=-4 \pi  G \frac{1}{\rho^2}\Big( \rho^4  (\pa_\rho\phi)^2+ (\pa_\tau\phi)^2\Big)-2\frac{\delta Q^{\rmm BR}}{Q}~,\label{Eq:BRpert00}\\
&\rho\pa_{\tau}\pa_{\rho}K-\pa_{\tau}K= - 8 \pi  G\rho \pa_\rho\phi \pa_\tau\phi~,\label{Eq:BRpert01}\\
&\pa^2_{\tau}K-\rho^3\pa_{\rho}K+\rho^2K-2 \rho^2\frac{\delta Q^{\rmm BR}}{Q}=-4 \pi  G \Big(\rho^4 (\pa_\rho\phi)^2+ (\pa_\tau\phi)^2\Big)~.\label{Eq:BRpert11}
\end{align}
\end{subequations}
The angular component of the perturbed field equations, Eq.~\eqref{Eq:leftovereqBR}, reads
\be\label{Eq:BRpertAngular}
\begin{split}
\pa_{\rho}\lf(\rho \pa_{\tau}H_{\tau\rho}\rg)+3K\rho -\frac{1}{2}\lf[\rho\pa_{\rho}^2H_{\tau\tau}-\pa_{\rho}H_{\tau\tau}+\rho\pa_{\tau}^2H_{\rho\rho}+\rho^4\pa_\rho H_{\rho\rho}+4\rho^3H_{\rho\rho}\rg]=\\
\frac{4\pi G}{\rho}\Big(\lf(\pa_{\tau}\phi\rg)^2-\rho^4\lf(\pa_{\rho}\phi\rg)^2 \Big)
+4\rho\frac{\delta Q^{\rmm BR}}{Q}~.
\end{split}
\ee

It is worth noting that combining Eqs.~\eqref{Eq:BRpert00}, \eqref{Eq:BRpert11}, we obtain
\be\label{Eq:KGeqKpertBR}
-\pa_{\tau}^2K+\rho^4\pa_{\rho}^2K+2\rho^3\pa_{\rho}K-2\rho^2K=-4\rho^2\frac{\delta Q^{\rmm BR}}{Q}~.
\ee
Eqs.~\eqref{EQ:BRpertAll} and \eqref{Eq:KGeqKpertBR} are familiar in the context of Jackiw-Teitelboim (JT) theory. Indeed, if we turn off the zero mode solution $K=\Phi_0\propto\delta Q$ in Eq.~\eqref{Eq:KGeqKpertBR}, then it reduces to the massive Klein-Gordon equation $(\Box_2-2/r_0^2)\Phi_{JT}=0$ on $AdS_2$ with $K=\Phi_{JT}$~. Moreover, the latter is simply the traceless part
of the JT dilaton equation of motion $g_{AB}\Box_2\Phi_{JT}-\nabla_A\nabla_B\Phi_{JT}-g_{AB}\Phi_{JT}=\nabla_A\phi\nabla_B\phi-{1\over 2}g_{AB}\nabla_A\phi\nabla^A\phi$~.

The solution to Eq.~\eqref{Eq:BRpert00} for the metric perturbation $K$ is
\be\label{Eq:SolutionBR}
K(\tau,\rho)=\frac{2\delta Q^{\rmm BR}}{Q}+8\pi G\, \chi(\tau)\rho-4\pi G\,\frac{\eta(\tau) }{\rho}- 4\pi G\rho \int_{\delta}^{\rho}\de \rho^{\prime}\;\frac{1}{\rho^{\prime\,3}} \int_{\delta}^{\rho^{\prime}}\de \rho^{\prime\prime}\; \frac{1}{\rho^{\prime\prime\,2}}\Big((\pa_\tau\phi)^2+\rho^{\prime\prime\,4}(\pa_{\rho^{\prime\prime}}\phi)^2 \Big) ~,
\ee
where $\chi(\tau)$ and $\eta(\tau)$ are subject to the following equations, obtained upon substituting \eqref{Eq:SolutionBR} into Eqs.~\eqref{Eq:BRpert01} and \eqref{Eq:BRpert11}
\be\label{Eq:ContinuityEqBR1}
\frac{\de\eta}{\de\tau}+\rho^2\pa_{\tau}\phi\pa_{\rho}\phi=\frac{\pa}{\pa \tau}\int_{\delta}^\rho \frac{\de\rho^{\prime}}{2\rho^{\prime\,2}}\Big( \lf(\pa_{\tau}\phi\rg)^2+ \rho^{\prime\,4}\lf(\pa_{\rho^{\prime}}\phi\rg)^2 \Big)~,
\ee
\be\label{Eq:ContinuityEqBR2}
\begin{split}
&\frac{\de^2\chi}{\de\tau^2}-\eta+\frac{1}{2\rho^2}\pa_\tau\lf(\rho^2\pa_\tau\phi \pa_\rho\phi \rg)+\frac{1}{2\rho}\Big( \lf(\pa_{\tau}\phi\rg)^2+ \rho^{4}\lf(\pa_{\rho}\phi\rg)^2 \Big)+\int_{\delta}^\rho \frac{\de\rho^{\prime}}{2\rho^{\prime\,2}}\Big( \lf(\pa_{\tau}\phi\rg)^2+ \rho^{\prime\,4}\lf(\pa_{\rho^{\prime}}\phi\rg)^2 \Big)\\
&-\frac{1}{2\rho^2}\pa_{\tau}^2\int_{\delta}^\rho \frac{\de\rho^{\prime}}{2\rho^{\prime\,2}}\Big( \lf(\pa_{\tau}\phi\rg)^2+ \rho^{\prime\,4}\lf(\pa_{\rho^{\prime}}\phi\rg)^2 \Big)-\pa_{\tau}^2\int_{\delta}^{\rho}\frac{\de\rho^{\prime}}{\rho^{\prime\,3}} \int_{\delta}^{\rho^{\prime}}\frac{\de\rho^{\prime\prime}}{2\rho^{\prime\prime\,2}}\Big( \lf(\pa_{\tau}\phi\rg)^2+ \rho^{\prime\prime\,4}\lf(\pa_{\rho^{\prime\prime}}\phi\rg)^2 \Big)=0~.
\end{split}
\ee
Upon differentiation with respect to $\rho$ and use of the Klein-Gordon equation \eqref{Eq:ScalarWaveEquationBR}, it is straightforward to verify that $\eta$ and $\chi$ are indeed pure functions of $\tau$~. The above are continuity equations, which determine $\eta$ up to a constant and $\chi$ up to a quadratic function of $\tau$~.

The general solution for the metric perturbation $H_{\tau\rho}$ is obtained by solving Eq.~\eqref{Eq:BRpertAngular}
\be\label{Eq:BRsol_tr}
\begin{split}
H_{\tau\rho}(\tau,\rho)=&\; \psi(\rho)+\frac{\omega(\tau)}{\rho}-\frac{\delta Q^{\rmm BR}}{Q}\rho\tau+12\pi G  \int_0^{\tau}\de\tau^{\prime}\,\eta(\tau^{\prime}) -  8\pi G\,\rho^2\int_0^{\tau}\de\tau^{\prime}\,\chi(\tau^{\prime})\\
&+\Xi(\tau,\rho)-\frac{4\pi G}{\rho}\int_{0}^{\tau}\de\tau^{\prime}\int_{\delta}^{\rho}\de\rho^{\prime}\frac{\rho^{\prime\,4}(\pa_{\rho^{\prime}}\phi)^2-(\pa_{\tau^{\prime}}\phi)^2}{\rho^{\prime}} \\
&+\frac{12\pi G }{\rho}\int_{0}^{\tau}\de\tau^{\prime}\int_{\delta}^{\rho}\de\rho^{\prime}\rho^{\prime\,2}\int_{\delta}^{\rho^{\prime}}\frac{\de\rho^{\prime\prime}}{\rho^{\prime\prime\,3}}\int_{\delta}^{\rho^{\prime\prime}}\de\rho^{\prime\prime\prime}\frac{(\pa_{\tau^{\prime}}\phi)^2+ \rho^{\prime\prime\prime\,4}(\pa_{\rho^{\prime\prime\prime}}\phi)^2}{\rho^{\prime\prime\prime\,2}}~,
\end{split}
\ee
where $\psi(\rho)$~, $\omega(\tau)$ are arbitrary functions, and we defined
\be\label{Eq:DefZ}
\Xi(\tau,\rho)\equiv\frac{1}{2\rho}\int_0^{\tau}\de\tau^{\prime}\;\int\de\rho\; \lf(\rho\pa_{\rho}^2H_{\tau\tau}-\pa_{\rho}H_{\tau\tau}+\rho\pa_{\tau^{\prime}}^2H_{\rho\rho}+\rho^4\pa_\rho H_{\rho\rho}+4\rho^3H_{\rho\rho}\rg)~.
\ee

Note that in the electrovacuum case ($\phi=constant$), $\eta$ is constant and $\chi$ is a quadratic function of $\tau$~,
\be\label{Eq:CoeffsBRelectrovacuum}
8\pi G\, \chi(\tau)=a+b\,\tau+c\,\tau^2~,\quad 4\pi G\, \eta(\tau) =c~.
\ee
The ${\rm SL(2,\mathbb{R})}$ transformation properties of the electrovacuum solutions were studied in detail in Ref.~\cite{Hadar:2020kry}, where it was understood that these solutions are classified based on the value of the Casimir invariant $b^2-4ac$ which in turn characterizes the gravitational backreaction of these linear solutions onto BR as an anabasis that is building asymptotically flat regions of Reissner-Nordstr{\"o}m black holes.

Lastly, the perturbed Misner-Sharp mass obtained from Eqs.~\eqref{Eq:MisnerSharpDefinition}, \eqref{Eq:MisnerSharpPerturbation} is
\be\label{Eq:PerturbedMassBR}
m_{\rm MS}=\frac{r_0}{2G}\lf(1+\frac{K}{2}\rg)~.
\ee
We note that in BR this quantity is gauge-invariant. This is in contrast with the corresponding calculation in ERN, Eq.~\eqref{Eq:MisnerSharpERN}.

\section{Matching perturbed ERN/BR spacetimes}\label{Sec:Matching}

In this section we show how the well-known identification of the BR spacetime with the near-horizon region of the ERN spacetime can be extended beyond the background level to also include perturbations.

For the background spacetimes, it can be readily seen that, in the chosen coordinates, the near-horizon limit of the ERN metric \eqref{Eq:Background_ERN} matches the BR metric \eqref{Eq:Background_BR}. The corresponding solutions for the Maxwell fields match as well. This matching can been seen it two ways. First, one can simply consider the small $\rho\ll 1$ approximation in ERN, effectively replacing every instance of $1+\rho$ with $1$~, and then one sees immediately that it indeed reduces to BR. It is worth pondering, however, exactly why that is. For it is by no means guaranteed that keeping leading order approximations  in each metric component of a solution to the Einstein equations, in some coordinate limit, will produce another solution of the Einstein equation. On the other hand, it is guaranteed if, following Geroch~\cite{Geroch:1969ca}, we can define a one-parameter family of coordinates in which to write the ERN solution, such that taking the parameter to zero the limit is regular. From this point of view, the reason that taking the near-horizon limit in the ``pedestrian'' sense of simply approximating $1+\rho\approx 1$ worked just fine for us so far is that the coordinates of \eqref{Eq:Background_ERN} and \eqref{Eq:Background_BR} are already properly aligned: the required one-parameter family of coordinates one can use to take the limit of ERN formally is simply obtained by $\rho\to\lambda\rho~, \tau\to\tau/\lambda$ in  \eqref{Eq:Background_ERN}, so that following this up by $\lambda\to 0$ yields exactly \eqref{Eq:Background_BR}.

The above considerations imply that when attempting to match the perturbed spacetimes as well, one expects to need a renewed alignment of coordinates which will now work up to the perturbative order one is working in. Equivalently, if we keep the coordinates fixed to the choice that aligns the backgrounds well, then we expect that we will need to align the gauges in which the perturbations are written in and match between BR and near-horizon ERN perturbations in these gauges only. We will see that this is indeed the case for the perturbations found in Sections~\ref{Sec:pertERN} and \ref{Sec:pertBR}, namely we will show that we can match the BR perturbations to the ``pedestrian'' near-horizon approximations of the ERN perturbations when we chose the gauge for the perturbations appropriately.

In our perturbative scheme, at leading order we only have the Klein-Gordon scalar propagating on the background spacetimes as a test field, namely satisfying Eqs.~\eqref{Eq:ScalarWaveEquationBR} and Eq.~\eqref{Eq:ScalarWaveEquationERN} in BR and ERN, respectively. Evidently, near the horizon the ERN equation reduces to the BR one. Indeed, the most general solution of the BR wave equation \eqref{Eq:ScalarWaveEquationBR}, given by $\phi(\tau,\rho)=\phi_L(\tau-1/\rho)+\phi_R(\tau+1/\rho)$~, can be matched to the corresponding near-horizon solution of the ERN wave equation \eqref{Eq:ScalarWaveEquationERN}. This is reviewed in Appendix~\ref{Sec:MAEscalar} using the method of matched asymptotic expansions in frequency space.
The matching of the scalar ensures that various integrals involving $\phi$ and its derivatives, as they appear in the solutions for metric perturbations in ERN and BR often match, in the near-horizon limit, simply by virtue of the integrands matching.

Moving to the next order in perturbation theory, we need to match the perturbations of the metric and Maxwell fields as well. As discussed above, in order to obtain $\overline{g}_{\mu\nu}^{\rmm ERN}\approx\overline{g}_{\mu\nu}^{\rmm BR}~,~\overline{{\cal F}}^{\rmm ERN}\approx \overline{{\cal F}}^{\rmm BR}$ near the horizon, we may use the coordinates adapted for matching the backgrounds and then ensure that the gauges are chosen appropriately for the perturbative solutions to match as well, $H_{AB}^{\rmm ERN}\approx H_{AB}^{\rmm BR}$~,~$K^{\rmm ERN}\approx K^{\rmm BR}$~.
More precisely, using the results for the perturbations derived in Sections \ref{Sec:pertERN} and \ref{Sec:pertBR}, we can match as follows:
\begin{enumerate}[label=(\roman*)]
\item 
First, by comparing the respective continuity equations in ERN and BR,  \eqref{Eq:MomentumFlux} and \eqref{Eq:ContinuityEqBR1}, we see that, given the matching of the scalar in the near-horizon approximation, we have:
	\be\label{Eq:MomentumFlux_identify}
	 q(\tau)=\eta(\tau)~.
	\ee	
\item Next, we ensure the matching of $H_{\rho\rho}$ and $H_{\tau\tau}$ components. These are arbitrary in both ERN and BR solutions (i.e. pure gauge) but for matching in the near-horizon approximation we need to assume that the $\rho\ll 1$ asymptotics of $H_{\rho\rho}$~, $H_{\tau\tau}$ in ERN and BR must be such that all terms which are not sub-leading compared to the corresponding background metric component match. 
\item Moving to the angular components of the two perturbed spacetimes, we need to match $\overline{r^2}$ between ERN and BR. We also need to match the solutions for the perturbed Maxwell fields, \eqref{Eq:PertMaxwellERN} and \eqref{Eq:PertMaxwellBR}, which combined with the requirement on  $\overline{r^2}$, implies that $\delta Q^{\rmm ERN}=\delta Q^{\rmm BR}$~. Hence, we will drop the superscript in the following and use the notation $\delta Q$ in both spacetimes. The matching of $\overline{r^2}$ then requires matching the two solutions for $K$, given in Eqs.~\eqref{Eq:WsolERN} and \eqref{Eq:SolutionBR}. This yields
\be\label{Eq:matchingf}
f(\tau)=8\pi G\, \chi(\tau)
\ee
as well as the following gauge-requirement condition
\be
-\frac{1}{\rho}\frac{\delta Q}{Q}+\frac{\rho}{(1+\rho)^2} \int \de\rho\; \frac{\rho\, H_{\rho\rho}}{1+\rho}=\frac{2\delta Q}{Q}+ \mathcal{O}(\rho)~,\nonumber
\ee
which translates into
\be\label{Eq:DynamicalGaugeFixing1}
H_{\rho\rho}= -\frac{2\delta Q}{Q}\rho^{-4}+\mathcal{O}\lf(\rho^{-3}\rg)~.
\ee
A comment is in order regarding the condition \eqref{Eq:matchingf}. Recall that $\chi(\tau)$ was fixed, up to a quadratic function of $\tau$, by the continuity equation \eqref{Eq:ContinuityEqBR2} in BR, while $f(\tau)$ was arbitrary in ERN but linked to the choice of antiderivatives in the indefinite integrals in Eq.~\eqref{Eq:WsolERN}. Additionally, we note in passing, that a quadratic function of $\tau$ in $\chi(\tau)$ (or $f(\tau)$) is interpreted, in the context of higher-order gravitational backreaction, as an anabasis perturbation responsible for building a background Reissner-Nordstrom black hole \cite{Hadar:2020kry}.
\item Lastly, we need to match the $H_{\tau\rho}$ solutions in \eqref{Eq:ERNsol_tr_Simp} and \eqref{Eq:BRsol_tr}. This requires
\be\label{Eq:GaugeCond3}
 s(\rho)=\psi(\rho)~,\quad \int_0^{\tau}\de\tau^{\prime}\, j(\tau^{\prime})=\omega(\tau)~,
\ee
as well as the following condition
\be\label{Eq:GaugeCond4}
\begin{split}
&-\frac{4\pi G (1+\rho)}{\rho}\int_{0}^{\tau}\de\tau^{\prime}\int_{\delta}^{\rho}\de\rho^{\prime}\frac{\rho^{\prime}}{(1+\rho^{\prime})^4}\int_{\delta}^{\rho^{\prime}}\de\rho^{\prime\prime}\frac{(1+\rho^{\prime\prime})^4(\pa_{\tau^{\prime}}\phi)^2+\rho^{\prime\prime\;4}(\pa_{\rho^{\prime\prime}}\phi)^2}{\rho^{\prime\prime\; 2}}\\
 &\hspace{8em} -\frac{1}{\rho(1+\rho)^2}\frac{\delta Q}{Q}\tau +\frac{1+\rho}{\rho}\int_0^{\tau}\de\tau^{\prime} \int\de\rho\,  Z(\tau^{\prime},\rho)\approx
 \Xi(\tau,\rho)-\frac{\delta Q}{Q}\rho\tau ~.
\end{split}
\ee
Recalling the definitions of $ Z$~\eqref{Eq:DefXtripletilde}, $\Xi$~\eqref{Eq:DefZ} and neglecting higher-order terms in the expansion of the various pre-factors, this condition boils down to
\be
\begin{split}
& \frac{1}{\rho}\int_0^{\tau}\de\tau^{\prime}\int\de\rho\, \lf[2H_{\tau\tau}-3\rho^2\int\de\rho\, \rho H_{\rho\rho}\rg]\approx\\
&\hspace{4em}\frac{4\pi G}{\rho}\int_{0}^{\tau}\de\tau^{\prime}\int_{\delta}^{\rho}\de\rho^{\prime}\rho^{\prime}\int_{\delta}^{\rho^{\prime}}\de\rho^{\prime\prime}\frac{(\pa_{\tau^{\prime}}\phi)^2+\rho^{\prime\prime\;4}(\pa_{\rho^{\prime\prime}}\phi)^2}{\rho^{\prime\prime\; 2}}
-\frac{\delta Q}{Q}\rho\tau +\frac{1}{\rho}\frac{\delta Q}{Q}\tau~,
\end{split}
\ee
which, after some straightforward algebra and differentiation, yields
\be\label{Eq:DynamicalGaugeFixing2}
2H_{\tau\tau}-3\rho^2\int\de\rho\, \rho H_{\rho\rho}\approx
4\pi G\rho\int_{\delta}^{\rho}\de\rho^{\prime}\frac{(\pa_{\tau}\phi)^2+\rho^{\prime\;4}(\pa_{\rho^{\prime}}\phi)^2}{\rho^{\prime\; 2}}-2\frac{\delta Q}{Q}\rho ~.
\ee
This is a further gauge-requirement condition. Notice that, unsurprisingly, the required choice of gauge in the end depends on the particular functional profile for the solution $\phi$~. 
For the matching in the electrovacuum case, see Appendix~\ref{Appendix:Electrovacuum}.
\end{enumerate}

Notice that, generically whenever $\delta Q\neq 0$~, the conditions on the required choice of gauge given above, specifically Eqs.~\eqref{Eq:DynamicalGaugeFixing1}, \eqref{Eq:DynamicalGaugeFixing2}, imply that the gauge is completely fixed at the necessary leading orders. On the other hand, if $\delta Q=0$ there is still gauge freedom left in the near-horizon limit and $H_{\rho\rho}$ is unconstrained.

\section{Conclusion}\label{Sec:Conclusion}

In this work we obtained general solutions for the evolution of linear spherically symmetric perturbations in the extremal Reissner-Nordstr{\"o}m  (ERN) and Bertotti-Robinson (BR) spacetimes in the presence of a massless scalar field. We derived precise conditions for matching the two perturbed spacetimes in the near-horizon region by means of suitable identifications and gauge adjustments. Generically, that is to say for a non-zero perturbation of the electric charge and a general scalar field profile, the matching can only be achieved in a specific gauge. Our results also show that in the special case where the electric charge is unperturbed, part of the gauge freedom survives the matching procedure.

The linearization of the ERN background around BR clearly has the same structure as some of the terms contributing to the solution for perturbations in BR. From a BR perspective alone it is not possible to disentangle the two types of perturbations at the approximation order we worked in. Higher-order perturbations are necessary in order to reconstruct the full perturbed ERN solution from BR and obtain the correct boundary conditions satisfied by BR perturbations. These should be the boundary conditions of a connected BR \cite{Hadar:2020kry}, as opposed to a stand-alone one, which doesn't have a well-behaved backreaction problem.

Further work is also required to understand the evolution of the apparent horizon of a perturbed extremal black hole in response to infalling matter fields. Although first-order perturbation theory, combined with matched asymptotic expansions, proved sufficient to study the evolution of the trapping horizon analytically in the case of a Schwarzschild black hole \cite{deCesare:2023rmg}, the generalization to the extremal case is not straightforward. In fact, we observe that the equation $2G\, m_{\rm MS}=r$~, which locates the apparent horizon in a generic spherically symmetric spacetime, reduces to an identity in the BR case, as can be easily realized using Eq.~\eqref{Eq:PerturbedMassBR}. Second-order perturbations break this identity, due to terms proportional to $\nabla_A K \nabla^A K$ that appear in the expansion of $m_{\rm MS}$ around BR.  An alternative definition of the mass of a perturbed extremal black hole, using near-horizon information only, was proposed in Ref.~\cite{Hadar:2020kry}, based on the Casimir invariant of the ${\rm SL(2,\mathbb{R})}$ symmetry group of BR. However, additional work is needed to clarify the connection with other definitions of the mass, and thus relate it to the apparent horizon. 

The present study can be generalized in several directions. For example, our methods can be applied to study the dynamics of perturbations with higher multipole numbers and non-zero spin as well as setups with AdS or dS far region asymptotics. 
More interestingly, instead of a neutral real scalar, one may consider a charged complex scalar field. This makes the dynamics richer and turns the charge perturbation $\delta Q$~, which is a constant in this work, into a dynamical quantity.

It would also be interesting to complement the analytical results obtained in our work with a numerical study of the connection problem, extending previous work on the interactions of scalar fields with black holes and their effects on black hole binaries~\cite{Traykova:2023qyv,Aurrekoetxea:2023jwk,Aurrekoetxea:2024cqd}.

A similar analysis can also be performed to study the impact of an extended gravitational sector on the ERN/BR matching problem in theories beyond general relativity. For instance, the formalism of Refs.~\cite{Gerlach:1979rw,Gerlach:1980tx,Gundlach:1999bt} has been recently extended in Ref.~\cite{Brizuela:2024smr} to bimetric gravity \cite{Hassan:2011zd}, a consistent nonlinear theory describing the interactions of massless and massive gravitons. 

Extensions to the axisymmetric case of extremal rotating black holes, for which ERN represents a useful toy model in spherical symmetry, can potentially open up the window to phenomenology of extremal rotating black holes. Indeed, the near-horizon region of extreme Kerr black holes with electromagnetic and matter fields is of interest for astrophysical applications.
As an example, Refs.~\cite{Camilloni:2020hns,Camilloni:2020qah} introduced an approach for constructing force-free magnetospheres in extreme Kerr black holes. Starting from force-free magnetospheres in the near-horizon region of such black holes, perturbative techniques are applied to extend solution beyond the near-horizon region. The methods developed in the present paper can potentially provide further insights into the (implicit) gauge adjustments that are necessary when constructing solutions outside the near-horizon region. Indeed, as for gravitational perturbations, our techniques could similarly help illuminate the process of gauge adjustments required to achieve consistency in matching gauge and field perturbations.

\section*{Acknowledgements}
We would like to thank Shahar Hadar and Eric Gourgoulhon for helpful comments.
This research is implemented in the framework of H.F.R.I. call ``Basic research Financing (Horizontal support of all Sciences)'' under the National Recovery and Resilience Plan ``Greece 2.0'' funded by the European Union --- NextGenerationEU (H.F.R.I. Project Number: 15384). APP is also partially supported by the European MSCA grant HORIZON-MSCA-2022-PF-01-01.
 MdC acknowledges support from the INFN Iniziativa Specifica GeoSymQFT. The work of RO is supported by the R{\'e}gion {\^I}le-de-France within the DIM ACAV+ project SYMONGRAV (Sym{\'e}tries asymptotiques et ondes gravitationnelles). Some of our calculations were performed with Mathematica packages of the {\it xAct} project \cite{xact}, particularly {\it xPert} \cite{Brizuela:2008ra}, as well as {\it RGTC} \cite{rgtc}.

\appendix

\section{The connection problem for the scalar}\label{Sec:MAEscalar}

\renewcommand{\theequation}{A.\arabic{equation}}

The connection problem for a massless Klein-Gordon scalar on fixed BR and ERN backgrounds is easy to solve completely using the method of matched asymptotic expansions in frequency space.

The most general spherically symmetric solution of the BR wave equation \eqref{Eq:ScalarWaveEquationBR} is given by $\phi(\tau,\rho)=\phi_L(\tau-1/\rho)+\phi_R(\tau+1/\rho)$ for arbitrary functions $\phi_L\,, \phi_R$~. In frequency space, expanding in modes $\phi\sim e^{-i\omega \tau}\phi_\omega(\rho)$~, this means
\be\label{Eq:ScalarWaveBRsolution}
\phi_\omega=C_1 e^{i\omega/\rho}+C_2 e^{-i\omega/\rho}
\ee
for arbitrary $C_1\,, C_2$~. Any such solution can be matched to the low-energy near-horizon modes of the ERN wave equation \eqref{Eq:ScalarWaveEquationERN} as follows. 

Expanding in modes $\phi\sim e^{-i\omega \tau}\phi_\omega(\rho)$ and changing variable to $\phi_\omega(\rho)=\rho^2 \psi_\omega(\rho)$~, Eq.~\eqref{Eq:ScalarWaveEquationERN} becomes
\be\label{Eq:ScalarWaveEquationERNfourier}
\rho^4 \psi_\omega^{\prime\prime}+6\rho^3 \psi_\omega^{\prime}+\left[6\rho^2(1+\omega^2)+\omega^2(1+4\rho+4\rho^3+\rho^4)\right]\psi_\omega=0~.
\ee
Consider low-energy modes, $\omega\ll 1$~, of this equation and divide the spacetime in three regions
\begin{subequations}
\begin{align}
	&\textrm{\it near:}\qquad\qquad\quad \,\rho\ll 1~,\label{near region}\\
	&\textrm{\it static:}\qquad\, \omega\ll \,\rho\ll 1/\omega~,\label{static region}\\
	&\textrm{\it far:}\qquad\quad\,\, 1\ll\,\rho~. \label{far region}
\end{align}
\end{subequations}
The {\it near } region might as well be called the BR region since here Eq.~\eqref{Eq:ScalarWaveEquationERNfourier} becomes
\be
\rho^4 \psi_\omega^{\prime\prime}+6\rho^3 \psi_\omega^{\prime}+\left(6\rho^2+\omega^2\right)\psi_\omega=0~,
\ee
which is identical to the mode expansion of the BR equation \eqref{Eq:ScalarWaveEquationBR} and has the solution corresponding to Eq.~\eqref{Eq:ScalarWaveBRsolution} above:
\be\label{Eq:NearSoln}
\psi_\omega^{\rmm near}={1\over\rho^2}(C_1^{\rmm near} e^{i\omega/\rho}+C_2^{\rmm near} e^{-i\omega/\rho})~.
\ee
For $\omega\ll1$~, the intermediate {\it static} region is overlapping with both the {\it near} and the {\it far} regions, and may therefore be used to connect the {\it near} and {\it far} solutions. It is called static because here Eq.~\eqref{Eq:ScalarWaveEquationERNfourier} becomes $\omega$-independent
\be
\rho^2 \psi_\omega^{\prime\prime}+6\rho \psi_\omega^{\prime}+6\psi_\omega=0~,
\ee
with solution
\be\label{Eq:StaticSoln}
\psi_\omega^{\rmm static}=\frac{C_1^{\rmm static}}{\rho^2}+\frac{C_2^{\rmm static}}{\rho^3}~.
\ee
Finally, in the {\it far} region Eq.~\eqref{Eq:ScalarWaveEquationERNfourier} becomes
\be
\rho^2 \psi_\omega^{\prime\prime}+6\rho \psi_\omega^{\prime}+\left(6+\rho^2\omega^2\right)\psi_\omega=0~,
\ee
with solution
\be\label{Eq:FarSoln}
\psi_\omega^{\rmm far}={1\over\rho^3}(C_1^{\rmm far} e^{i\omega\rho}+C_2^{\rmm far} e^{-i\omega\rho})~.
\ee
Matching $\psi_\omega^{\rmm static}$ with $\psi_\omega^{\rmm near}$ in the {\it near-static} overlap region $\omega\ll \rho\ll 1$ we get $C_1^{\rmm static}=C_1^{\rmm near}+C_2^{\rmm near}$ and $C_2^{\rmm static}=i\omega(C_1^{\rmm near}-C_2^{\rmm near})$~. 
Similarly, matching $\psi_\omega^{\rmm static}$ with $\psi_\omega^{\rmm far}$ in the {\it far-static} overlap region $1\ll \rho\ll 1/\omega$ we get $C_1^{\rmm static}=i\omega(C_1^{\rmm far}-C_2^{\rmm far})$ and $C_2^{\rmm static}=C_1^{\rmm far}+C_2^{\rmm far}$~. As a result, any near-horizon BR solution \eqref{Eq:NearSoln} uniquely matches onto the low-energy ERN solution with {\it far} behavior given by \eqref{Eq:FarSoln} with
\begin{subequations}
\begin{align}
C_1^{\rmm far}+C_2^{\rmm far}&=i\omega(C_1^{\rmm near}-C_2^{\rmm near})~,\\
i\omega(C_1^{\rmm far}-C_2^{\rmm far})&=C_1^{\rmm near}+C_2^{\rmm near}~.
\end{align}
\end{subequations}

\section{Matching electrovacuum perturbations}\label{Appendix:Electrovacuum}

\renewcommand{\theequation}{B.\arabic{equation}}

The solutions for electrovacuum perturbations in ERN and BR may be obtained by setting $\phi=constant$ in the general solutions from Sections~\ref{Sec:pertERN} and \ref{Sec:pertBR}, specifically Eqs.~\eqref{Eq:WsolERN}, \eqref{Eq:ERNsol_tr_Simp} and \eqref{Eq:SolutionBR}, \eqref{Eq:BRsol_tr}, respectively. Indeed, the electrovacuum perturbations of ERN are simply the two-parameter family of solutions labeled by $k$ and $\delta Q$, while the corresponding BR solution is the four-parameter family labeled by $(a,b,c)$ and $\delta Q$.
The matching of perturbations of the metric and Maxwell field in the two spacetimes proceeds as in the general case in Section~\ref{Sec:Matching}, with minor differences. 
Naturally, several simplifications occur and we highlight those below.

First, Eq.~\eqref{Eq:MomentumFlux_identify} is replaced by
\be
k=\frac{c}{4\pi G}~,
\ee	
which simply identifies the integration constants arising from the continuity equations.
Similarly, Eq.~\eqref{Eq:matchingf} is replaced by
\be
f(\tau)=8\pi G\, \chi(\tau)=a+b\tau+c\tau^2~,
\ee
which again fixes the arbitrary (gauge) function in ERN in terms of the parameters of the BR solution.
On the other hand, the gauge conditions \eqref{Eq:DynamicalGaugeFixing1} and \eqref{Eq:GaugeCond3} are not modified.
Eq.~\eqref{Eq:GaugeCond4} simplifies to
\be
-\frac{1}{\rho(1+\rho)^2}\frac{\delta Q}{Q}\tau +\frac{1+\rho}{\rho}\int_0^{\tau}\de\tau^{\prime} \int\de\rho\,  Z(\tau^{\prime},\rho)\approx
 \Xi(\tau,\rho)-\frac{\delta Q}{Q}\rho\tau ~,
\ee
which, following entirely analogous steps to those leading to Eq.~\eqref{Eq:DynamicalGaugeFixing2}, boils down to
\be
2H_{\tau\tau}-3\rho^2\int\de\rho\, \rho H_{\rho\rho}\approx
-2\frac{\delta Q}{Q}\rho ~.
\ee

\bibliography{ERN-BR_matching}

\begin{thebibliography}{35}%
\makeatletter
\providecommand \@ifxundefined [1]{%
 \@ifx{#1\undefined}
}%
\providecommand \@ifnum [1]{%
 \ifnum #1\expandafter \@firstoftwo
 \else \expandafter \@secondoftwo
 \fi
}%
\providecommand \@ifx [1]{%
 \ifx #1\expandafter \@firstoftwo
 \else \expandafter \@secondoftwo
 \fi
}%
\providecommand \natexlab [1]{#1}%
\providecommand \enquote  [1]{``#1''}%
\providecommand \bibnamefont  [1]{#1}%
\providecommand \bibfnamefont [1]{#1}%
\providecommand \citenamefont [1]{#1}%
\providecommand \href@noop [0]{\@secondoftwo}%
\providecommand \href [0]{\begingroup \@sanitize@url \@href}%
\providecommand \@href[1]{\@@startlink{#1}\@@href}%
\providecommand \@@href[1]{\endgroup#1\@@endlink}%
\providecommand \@sanitize@url [0]{\catcode `\\12\catcode `\$12\catcode
  `\&12\catcode `\#12\catcode `\^12\catcode `\_12\catcode `\%12\relax}%
\providecommand \@@startlink[1]{}%
\providecommand \@@endlink[0]{}%
\providecommand \url  [0]{\begingroup\@sanitize@url \@url }%
\providecommand \@url [1]{\endgroup\@href {#1}{\urlprefix }}%
\providecommand \urlprefix  [0]{URL }%
\providecommand \Eprint [0]{\href }%
\providecommand \doibase [0]{http://dx.doi.org/}%
\providecommand \selectlanguage [0]{\@gobble}%
\providecommand \bibinfo  [0]{\@secondoftwo}%
\providecommand \bibfield  [0]{\@secondoftwo}%
\providecommand \translation [1]{[#1]}%
\providecommand \BibitemOpen [0]{}%
\providecommand \bibitemStop [0]{}%
\providecommand \bibitemNoStop [0]{.\EOS\space}%
\providecommand \EOS [0]{\spacefactor3000\relax}%
\providecommand \BibitemShut  [1]{\csname bibitem#1\endcsname}%
\let\auto@bib@innerbib\@empty
\bibitem [{\citenamefont {Kunduri}\ \emph {et~al.}(2007)\citenamefont
  {Kunduri}, \citenamefont {Lucietti},\ and\ \citenamefont
  {Reall}}]{Kunduri:2007vf}%
  \BibitemOpen
  \bibfield  {author} {\bibinfo {author} {\bibfnamefont {Hari~K.}\ \bibnamefont
  {Kunduri}}, \bibinfo {author} {\bibfnamefont {James}\ \bibnamefont
  {Lucietti}}, \ and\ \bibinfo {author} {\bibfnamefont {Harvey~S.}\
  \bibnamefont {Reall}},\ }\bibfield  {title} {\enquote {\bibinfo {title}
  {{Near-horizon symmetries of extremal black holes}},}\ }\href {\doibase
  10.1088/0264-9381/24/16/012} {\bibfield  {journal} {\bibinfo  {journal}
  {Class. Quant. Grav.}\ }\textbf {\bibinfo {volume} {24}},\ \bibinfo {pages}
  {4169--4190} (\bibinfo {year} {2007})},\ \Eprint
  {http://arxiv.org/abs/0705.4214} {arXiv:0705.4214 [hep-th]} \BibitemShut
  {NoStop}%
\bibitem [{\citenamefont {Carter}(2009)}]{Carter2009}%
  \BibitemOpen
  \bibfield  {author} {\bibinfo {author} {\bibfnamefont {Brandon}\ \bibnamefont
  {Carter}},\ }\bibfield  {title} {\enquote {\bibinfo {title} {Republication
  of: Black hole equilibrium states},}\ }\href {\doibase
  10.1007/s10714-009-0888-5} {\bibfield  {journal} {\bibinfo  {journal}
  {General Relativity and Gravitation}\ }\textbf {\bibinfo {volume} {41}},\
  \bibinfo {pages} {2873--2938} (\bibinfo {year} {2009})}\BibitemShut {NoStop}%
\bibitem [{\citenamefont {Porfyriadis}(2018)}]{Porfyriadis:2018yag}%
  \BibitemOpen
  \bibfield  {author} {\bibinfo {author} {\bibfnamefont {Achilleas~P.}\
  \bibnamefont {Porfyriadis}},\ }\bibfield  {title} {\enquote {\bibinfo {title}
  {{Scattering of gravitational and electromagnetic waves off $AdS_2\times S^2$
  in extreme Reissner-Nordstrom}},}\ }\href {\doibase 10.1007/JHEP07(2018)064}
  {\bibfield  {journal} {\bibinfo  {journal} {JHEP}\ }\textbf {\bibinfo
  {volume} {07}},\ \bibinfo {pages} {064} (\bibinfo {year} {2018})},\ \Eprint
  {http://arxiv.org/abs/1805.12409} {arXiv:1805.12409 [hep-th]} \BibitemShut
  {NoStop}%
\bibitem [{\citenamefont {Porfyriadis}(2019)}]{Porfyriadis:2018jlw}%
  \BibitemOpen
  \bibfield  {author} {\bibinfo {author} {\bibfnamefont {Achilleas~P.}\
  \bibnamefont {Porfyriadis}},\ }\bibfield  {title} {\enquote {\bibinfo {title}
  {{Near- $AdS_2$ perturbations and the connection with near-extreme
  Reissner\textendash{}Nordstrom}},}\ }\href {\doibase
  10.1140/epjc/s10052-019-7347-6} {\bibfield  {journal} {\bibinfo  {journal}
  {Eur. Phys. J. C}\ }\textbf {\bibinfo {volume} {79}},\ \bibinfo {pages} {841}
  (\bibinfo {year} {2019})},\ \Eprint {http://arxiv.org/abs/1806.07097}
  {arXiv:1806.07097 [hep-th]} \BibitemShut {NoStop}%
\bibitem [{\citenamefont {Maldacena}\ \emph {et~al.}(1999)\citenamefont
  {Maldacena}, \citenamefont {Michelson},\ and\ \citenamefont
  {Strominger}}]{Maldacena:1998uz}%
  \BibitemOpen
  \bibfield  {author} {\bibinfo {author} {\bibfnamefont {Juan~Martin}\
  \bibnamefont {Maldacena}}, \bibinfo {author} {\bibfnamefont {Jeremy}\
  \bibnamefont {Michelson}}, \ and\ \bibinfo {author} {\bibfnamefont {Andrew}\
  \bibnamefont {Strominger}},\ }\bibfield  {title} {\enquote {\bibinfo {title}
  {{Anti-de Sitter fragmentation}},}\ }\href {\doibase
  10.1088/1126-6708/1999/02/011} {\bibfield  {journal} {\bibinfo  {journal}
  {JHEP}\ }\textbf {\bibinfo {volume} {02}},\ \bibinfo {pages} {011} (\bibinfo
  {year} {1999})},\ \Eprint {http://arxiv.org/abs/hep-th/9812073}
  {arXiv:hep-th/9812073} \BibitemShut {NoStop}%
\bibitem [{\citenamefont {Almheiri}\ and\ \citenamefont
  {Polchinski}(2015)}]{Almheiri:2014cka}%
  \BibitemOpen
  \bibfield  {author} {\bibinfo {author} {\bibfnamefont {Ahmed}\ \bibnamefont
  {Almheiri}}\ and\ \bibinfo {author} {\bibfnamefont {Joseph}\ \bibnamefont
  {Polchinski}},\ }\bibfield  {title} {\enquote {\bibinfo {title} {{Models of
  AdS$_{2}$ backreaction and holography}},}\ }\href {\doibase
  10.1007/JHEP11(2015)014} {\bibfield  {journal} {\bibinfo  {journal} {JHEP}\
  }\textbf {\bibinfo {volume} {11}},\ \bibinfo {pages} {014} (\bibinfo {year}
  {2015})},\ \Eprint {http://arxiv.org/abs/1402.6334} {arXiv:1402.6334
  [hep-th]} \BibitemShut {NoStop}%
\bibitem [{\citenamefont {Jensen}(2016)}]{Jensen:2016pah}%
  \BibitemOpen
  \bibfield  {author} {\bibinfo {author} {\bibfnamefont {Kristan}\ \bibnamefont
  {Jensen}},\ }\bibfield  {title} {\enquote {\bibinfo {title} {{Chaos in
  AdS$_2$ Holography}},}\ }\href {\doibase 10.1103/PhysRevLett.117.111601}
  {\bibfield  {journal} {\bibinfo  {journal} {Phys. Rev. Lett.}\ }\textbf
  {\bibinfo {volume} {117}},\ \bibinfo {pages} {111601} (\bibinfo {year}
  {2016})},\ \Eprint {http://arxiv.org/abs/1605.06098} {arXiv:1605.06098
  [hep-th]} \BibitemShut {NoStop}%
\bibitem [{\citenamefont {Maldacena}\ \emph {et~al.}(2016)\citenamefont
  {Maldacena}, \citenamefont {Stanford},\ and\ \citenamefont
  {Yang}}]{Maldacena:2016upp}%
  \BibitemOpen
  \bibfield  {author} {\bibinfo {author} {\bibfnamefont {Juan}\ \bibnamefont
  {Maldacena}}, \bibinfo {author} {\bibfnamefont {Douglas}\ \bibnamefont
  {Stanford}}, \ and\ \bibinfo {author} {\bibfnamefont {Zhenbin}\ \bibnamefont
  {Yang}},\ }\bibfield  {title} {\enquote {\bibinfo {title} {{Conformal
  symmetry and its breaking in two dimensional Nearly Anti-de-Sitter space}},}\
  }\href {\doibase 10.1093/ptep/ptw124} {\bibfield  {journal} {\bibinfo
  {journal} {PTEP}\ }\textbf {\bibinfo {volume} {2016}},\ \bibinfo {pages}
  {12C104} (\bibinfo {year} {2016})},\ \Eprint
  {http://arxiv.org/abs/1606.01857} {arXiv:1606.01857 [hep-th]} \BibitemShut
  {NoStop}%
\bibitem [{\citenamefont {Engels\"oy}\ \emph {et~al.}(2016)\citenamefont
  {Engels\"oy}, \citenamefont {Mertens},\ and\ \citenamefont
  {Verlinde}}]{Engelsoy:2016xyb}%
  \BibitemOpen
  \bibfield  {author} {\bibinfo {author} {\bibfnamefont {Julius}\ \bibnamefont
  {Engels\"oy}}, \bibinfo {author} {\bibfnamefont {Thomas~G.}\ \bibnamefont
  {Mertens}}, \ and\ \bibinfo {author} {\bibfnamefont {Herman}\ \bibnamefont
  {Verlinde}},\ }\bibfield  {title} {\enquote {\bibinfo {title} {{An
  investigation of AdS$_{2}$ backreaction and holography}},}\ }\href {\doibase
  10.1007/JHEP07(2016)139} {\bibfield  {journal} {\bibinfo  {journal} {JHEP}\
  }\textbf {\bibinfo {volume} {07}},\ \bibinfo {pages} {139} (\bibinfo {year}
  {2016})},\ \Eprint {http://arxiv.org/abs/1606.03438} {arXiv:1606.03438
  [hep-th]} \BibitemShut {NoStop}%
\bibitem [{\citenamefont {Mertens}\ and\ \citenamefont
  {Turiaci}(2023)}]{Mertens:2022irh}%
  \BibitemOpen
  \bibfield  {author} {\bibinfo {author} {\bibfnamefont {Thomas~G.}\
  \bibnamefont {Mertens}}\ and\ \bibinfo {author} {\bibfnamefont {Gustavo~J.}\
  \bibnamefont {Turiaci}},\ }\bibfield  {title} {\enquote {\bibinfo {title}
  {{Solvable models of quantum black holes: a review on
  Jackiw\textendash{}Teitelboim gravity}},}\ }\href {\doibase
  10.1007/s41114-023-00046-1} {\bibfield  {journal} {\bibinfo  {journal}
  {Living Rev. Rel.}\ }\textbf {\bibinfo {volume} {26}},\ \bibinfo {pages} {4}
  (\bibinfo {year} {2023})},\ \Eprint {http://arxiv.org/abs/2210.10846}
  {arXiv:2210.10846 [hep-th]} \BibitemShut {NoStop}%
\bibitem [{\citenamefont {S\'arosi}(2018)}]{Sarosi:2017ykf}%
  \BibitemOpen
  \bibfield  {author} {\bibinfo {author} {\bibfnamefont {G\'abor}\ \bibnamefont
  {S\'arosi}},\ }\bibfield  {title} {\enquote {\bibinfo {title} {{AdS$_{2}$
  holography and the SYK model}},}\ }\href {\doibase 10.22323/1.323.0001}
  {\bibfield  {journal} {\bibinfo  {journal} {PoS}\ }\textbf {\bibinfo {volume}
  {Modave2017}},\ \bibinfo {pages} {001} (\bibinfo {year} {2018})},\ \Eprint
  {http://arxiv.org/abs/1711.08482} {arXiv:1711.08482 [hep-th]} \BibitemShut
  {NoStop}%
\bibitem [{\citenamefont {Teitelboim}(1983)}]{Teitelboim:1983ux}%
  \BibitemOpen
  \bibfield  {author} {\bibinfo {author} {\bibfnamefont {C.}~\bibnamefont
  {Teitelboim}},\ }\bibfield  {title} {\enquote {\bibinfo {title} {{Gravitation
  and Hamiltonian Structure in Two Space-Time Dimensions}},}\ }\href {\doibase
  10.1016/0370-2693(83)90012-6} {\bibfield  {journal} {\bibinfo  {journal}
  {Phys. Lett. B}\ }\textbf {\bibinfo {volume} {126}},\ \bibinfo {pages}
  {41--45} (\bibinfo {year} {1983})}\BibitemShut {NoStop}%
\bibitem [{\citenamefont {Jackiw}(1985)}]{Jackiw:1984je}%
  \BibitemOpen
  \bibfield  {author} {\bibinfo {author} {\bibfnamefont {R.}~\bibnamefont
  {Jackiw}},\ }\bibfield  {title} {\enquote {\bibinfo {title} {{Lower
  Dimensional Gravity}},}\ }\href {\doibase 10.1016/0550-3213(85)90448-1}
  {\bibfield  {journal} {\bibinfo  {journal} {Nucl. Phys. B}\ }\textbf
  {\bibinfo {volume} {252}},\ \bibinfo {pages} {343--356} (\bibinfo {year}
  {1985})}\BibitemShut {NoStop}%
\bibitem [{\citenamefont {Almheiri}\ and\ \citenamefont
  {Kang}(2016)}]{Almheiri:2016fws}%
  \BibitemOpen
  \bibfield  {author} {\bibinfo {author} {\bibfnamefont {Ahmed}\ \bibnamefont
  {Almheiri}}\ and\ \bibinfo {author} {\bibfnamefont {Byungwoo}\ \bibnamefont
  {Kang}},\ }\bibfield  {title} {\enquote {\bibinfo {title} {{Conformal
  Symmetry Breaking and Thermodynamics of Near-Extremal Black Holes}},}\ }\href
  {\doibase 10.1007/JHEP10(2016)052} {\bibfield  {journal} {\bibinfo  {journal}
  {JHEP}\ }\textbf {\bibinfo {volume} {10}},\ \bibinfo {pages} {052} (\bibinfo
  {year} {2016})},\ \Eprint {http://arxiv.org/abs/1606.04108} {arXiv:1606.04108
  [hep-th]} \BibitemShut {NoStop}%
\bibitem [{\citenamefont {Nayak}\ \emph {et~al.}(2018)\citenamefont {Nayak},
  \citenamefont {Shukla}, \citenamefont {Soni}, \citenamefont {Trivedi},\ and\
  \citenamefont {Vishal}}]{Nayak:2018qej}%
  \BibitemOpen
  \bibfield  {author} {\bibinfo {author} {\bibfnamefont {Pranjal}\ \bibnamefont
  {Nayak}}, \bibinfo {author} {\bibfnamefont {Ashish}\ \bibnamefont {Shukla}},
  \bibinfo {author} {\bibfnamefont {Ronak~M.}\ \bibnamefont {Soni}}, \bibinfo
  {author} {\bibfnamefont {Sandip~P.}\ \bibnamefont {Trivedi}}, \ and\ \bibinfo
  {author} {\bibfnamefont {V.}~\bibnamefont {Vishal}},\ }\bibfield  {title}
  {\enquote {\bibinfo {title} {{On the Dynamics of Near-Extremal Black
  Holes}},}\ }\href {\doibase 10.1007/JHEP09(2018)048} {\bibfield  {journal}
  {\bibinfo  {journal} {JHEP}\ }\textbf {\bibinfo {volume} {09}},\ \bibinfo
  {pages} {048} (\bibinfo {year} {2018})},\ \Eprint
  {http://arxiv.org/abs/1802.09547} {arXiv:1802.09547 [hep-th]} \BibitemShut
  {NoStop}%
\bibitem [{\citenamefont {Moitra}\ \emph
  {et~al.}(2019{\natexlab{a}})\citenamefont {Moitra}, \citenamefont {Trivedi},\
  and\ \citenamefont {Vishal}}]{Moitra:2018jqs}%
  \BibitemOpen
  \bibfield  {author} {\bibinfo {author} {\bibfnamefont {Upamanyu}\
  \bibnamefont {Moitra}}, \bibinfo {author} {\bibfnamefont {Sandip~P.}\
  \bibnamefont {Trivedi}}, \ and\ \bibinfo {author} {\bibfnamefont
  {V.}~\bibnamefont {Vishal}},\ }\bibfield  {title} {\enquote {\bibinfo {title}
  {{Extremal and near-extremal black holes and near-CFT$_{1}$}},}\ }\href
  {\doibase 10.1007/JHEP07(2019)055} {\bibfield  {journal} {\bibinfo  {journal}
  {JHEP}\ }\textbf {\bibinfo {volume} {07}},\ \bibinfo {pages} {055} (\bibinfo
  {year} {2019}{\natexlab{a}})},\ \Eprint {http://arxiv.org/abs/1808.08239}
  {arXiv:1808.08239 [hep-th]} \BibitemShut {NoStop}%
\bibitem [{\citenamefont {Castro}\ \emph {et~al.}(2018)\citenamefont {Castro},
  \citenamefont {Larsen},\ and\ \citenamefont
  {Papadimitriou}}]{Castro:2018ffi}%
  \BibitemOpen
  \bibfield  {author} {\bibinfo {author} {\bibfnamefont {Alejandra}\
  \bibnamefont {Castro}}, \bibinfo {author} {\bibfnamefont {Finn}\ \bibnamefont
  {Larsen}}, \ and\ \bibinfo {author} {\bibfnamefont {Ioannis}\ \bibnamefont
  {Papadimitriou}},\ }\bibfield  {title} {\enquote {\bibinfo {title} {{5D
  rotating black holes and the nAdS$_{2}$/nCFT$_{1}$ correspondence}},}\ }\href
  {\doibase 10.1007/JHEP10(2018)042} {\bibfield  {journal} {\bibinfo  {journal}
  {JHEP}\ }\textbf {\bibinfo {volume} {10}},\ \bibinfo {pages} {042} (\bibinfo
  {year} {2018})},\ \Eprint {http://arxiv.org/abs/1807.06988} {arXiv:1807.06988
  [hep-th]} \BibitemShut {NoStop}%
\bibitem [{\citenamefont {Moitra}\ \emph
  {et~al.}(2019{\natexlab{b}})\citenamefont {Moitra}, \citenamefont {Sake},
  \citenamefont {Trivedi},\ and\ \citenamefont {Vishal}}]{Moitra:2019bub}%
  \BibitemOpen
  \bibfield  {author} {\bibinfo {author} {\bibfnamefont {Upamanyu}\
  \bibnamefont {Moitra}}, \bibinfo {author} {\bibfnamefont {Sunil~Kumar}\
  \bibnamefont {Sake}}, \bibinfo {author} {\bibfnamefont {Sandip~P.}\
  \bibnamefont {Trivedi}}, \ and\ \bibinfo {author} {\bibfnamefont
  {V.}~\bibnamefont {Vishal}},\ }\bibfield  {title} {\enquote {\bibinfo {title}
  {{Jackiw-Teitelboim Gravity and Rotating Black Holes}},}\ }\href {\doibase
  10.1007/JHEP11(2019)047} {\bibfield  {journal} {\bibinfo  {journal} {JHEP}\
  }\textbf {\bibinfo {volume} {11}},\ \bibinfo {pages} {047} (\bibinfo {year}
  {2019}{\natexlab{b}})},\ \Eprint {http://arxiv.org/abs/1905.10378}
  {arXiv:1905.10378 [hep-th]} \BibitemShut {NoStop}%
\bibitem [{\citenamefont {Castro}\ \emph {et~al.}(2021)\citenamefont {Castro},
  \citenamefont {Godet}, \citenamefont {Sim\'on}, \citenamefont {Song},\ and\
  \citenamefont {Yu}}]{Castro:2021csm}%
  \BibitemOpen
  \bibfield  {author} {\bibinfo {author} {\bibfnamefont {Alejandra}\
  \bibnamefont {Castro}}, \bibinfo {author} {\bibfnamefont {Victor}\
  \bibnamefont {Godet}}, \bibinfo {author} {\bibfnamefont {Joan}\ \bibnamefont
  {Sim\'on}}, \bibinfo {author} {\bibfnamefont {Wei}\ \bibnamefont {Song}}, \
  and\ \bibinfo {author} {\bibfnamefont {Boyang}\ \bibnamefont {Yu}},\
  }\bibfield  {title} {\enquote {\bibinfo {title} {{Gravitational perturbations
  from NHEK to Kerr}},}\ }\href {\doibase 10.1007/JHEP07(2021)218} {\bibfield
  {journal} {\bibinfo  {journal} {JHEP}\ }\textbf {\bibinfo {volume} {07}},\
  \bibinfo {pages} {218} (\bibinfo {year} {2021})},\ \Eprint
  {http://arxiv.org/abs/2102.08060} {arXiv:2102.08060 [hep-th]} \BibitemShut
  {NoStop}%
\bibitem [{\citenamefont {Hadar}\ \emph {et~al.}(2021)\citenamefont {Hadar},
  \citenamefont {Lupsasca},\ and\ \citenamefont {Porfyriadis}}]{Hadar:2020kry}%
  \BibitemOpen
  \bibfield  {author} {\bibinfo {author} {\bibfnamefont {Shahar}\ \bibnamefont
  {Hadar}}, \bibinfo {author} {\bibfnamefont {Alexandru}\ \bibnamefont
  {Lupsasca}}, \ and\ \bibinfo {author} {\bibfnamefont {Achilleas~P.}\
  \bibnamefont {Porfyriadis}},\ }\bibfield  {title} {\enquote {\bibinfo {title}
  {{Extreme Black Hole Anabasis}},}\ }\href {\doibase 10.1007/JHEP03(2021)223}
  {\bibfield  {journal} {\bibinfo  {journal} {JHEP}\ }\textbf {\bibinfo
  {volume} {03}},\ \bibinfo {pages} {223} (\bibinfo {year} {2021})},\ \Eprint
  {http://arxiv.org/abs/2012.06562} {arXiv:2012.06562 [hep-th]} \BibitemShut
  {NoStop}%
\bibitem [{\citenamefont {Gerlach}\ and\ \citenamefont
  {Sengupta}(1980)}]{Gerlach:1980tx}%
  \BibitemOpen
  \bibfield  {author} {\bibinfo {author} {\bibfnamefont {U.~H.}\ \bibnamefont
  {Gerlach}}\ and\ \bibinfo {author} {\bibfnamefont {U.~K.}\ \bibnamefont
  {Sengupta}},\ }\bibfield  {title} {\enquote {\bibinfo {title}
  {{Gauge-invariant coupled gravitational, acoustical, and electromagnetic
  modes on most general spherical space-times}},}\ }\href {\doibase
  10.1103/PhysRevD.22.1300} {\bibfield  {journal} {\bibinfo  {journal} {Phys.
  Rev. D}\ }\textbf {\bibinfo {volume} {22}},\ \bibinfo {pages} {1300--1312}
  (\bibinfo {year} {1980})}\BibitemShut {NoStop}%
\bibitem [{\citenamefont {Gerlach}\ and\ \citenamefont
  {Sengupta}(1979)}]{Gerlach:1979rw}%
  \BibitemOpen
  \bibfield  {author} {\bibinfo {author} {\bibfnamefont {U.~H.}\ \bibnamefont
  {Gerlach}}\ and\ \bibinfo {author} {\bibfnamefont {U.~K.}\ \bibnamefont
  {Sengupta}},\ }\bibfield  {title} {\enquote {\bibinfo {title}
  {{Gauge-invariant perturbations on most general spherically symmetric
  space-times}},}\ }\href {\doibase 10.1103/PhysRevD.19.2268} {\bibfield
  {journal} {\bibinfo  {journal} {Phys. Rev. D}\ }\textbf {\bibinfo {volume}
  {19}},\ \bibinfo {pages} {2268--2272} (\bibinfo {year} {1979})}\BibitemShut
  {NoStop}%
\bibitem [{\citenamefont {Gundlach}\ and\ \citenamefont
  {Martin-Garcia}(2000)}]{Gundlach:1999bt}%
  \BibitemOpen
  \bibfield  {author} {\bibinfo {author} {\bibfnamefont {Carsten}\ \bibnamefont
  {Gundlach}}\ and\ \bibinfo {author} {\bibfnamefont {Jose~M.}\ \bibnamefont
  {Martin-Garcia}},\ }\bibfield  {title} {\enquote {\bibinfo {title} {{Gauge
  invariant and coordinate independent perturbations of stellar collapse. 1.
  The Interior}},}\ }\href {\doibase 10.1103/PhysRevD.61.084024} {\bibfield
  {journal} {\bibinfo  {journal} {Phys. Rev. D}\ }\textbf {\bibinfo {volume}
  {61}},\ \bibinfo {pages} {084024} (\bibinfo {year} {2000})},\ \Eprint
  {http://arxiv.org/abs/gr-qc/9906068} {arXiv:gr-qc/9906068} \BibitemShut
  {NoStop}%
\bibitem [{\citenamefont {Geroch}(1969)}]{Geroch:1969ca}%
  \BibitemOpen
  \bibfield  {author} {\bibinfo {author} {\bibfnamefont {Robert~P.}\
  \bibnamefont {Geroch}},\ }\bibfield  {title} {\enquote {\bibinfo {title}
  {{Limits of spacetimes}},}\ }\href {\doibase 10.1007/BF01645486} {\bibfield
  {journal} {\bibinfo  {journal} {Commun. Math. Phys.}\ }\textbf {\bibinfo
  {volume} {13}},\ \bibinfo {pages} {180--193} (\bibinfo {year}
  {1969})}\BibitemShut {NoStop}%
\bibitem [{\citenamefont {de~Cesare}\ and\ \citenamefont
  {Oliveri}(2023)}]{deCesare:2023rmg}%
  \BibitemOpen
  \bibfield  {author} {\bibinfo {author} {\bibfnamefont {Marco}\ \bibnamefont
  {de~Cesare}}\ and\ \bibinfo {author} {\bibfnamefont {Roberto}\ \bibnamefont
  {Oliveri}},\ }\bibfield  {title} {\enquote {\bibinfo {title} {{Backreaction
  of scalar waves on black holes at low frequencies}},}\ }\href {\doibase
  10.1103/PhysRevD.108.044050} {\bibfield  {journal} {\bibinfo  {journal}
  {Phys. Rev. D}\ }\textbf {\bibinfo {volume} {108}},\ \bibinfo {pages}
  {044050} (\bibinfo {year} {2023})},\ \Eprint
  {http://arxiv.org/abs/2305.04970} {arXiv:2305.04970 [gr-qc]} \BibitemShut
  {NoStop}%
\bibitem [{\citenamefont {Traykova}\ \emph {et~al.}(2023)\citenamefont
  {Traykova}, \citenamefont {Vicente}, \citenamefont {Clough}, \citenamefont
  {Helfer}, \citenamefont {Berti}, \citenamefont {Ferreira},\ and\
  \citenamefont {Hui}}]{Traykova:2023qyv}%
  \BibitemOpen
  \bibfield  {author} {\bibinfo {author} {\bibfnamefont {Dina}\ \bibnamefont
  {Traykova}}, \bibinfo {author} {\bibfnamefont {Rodrigo}\ \bibnamefont
  {Vicente}}, \bibinfo {author} {\bibfnamefont {Katy}\ \bibnamefont {Clough}},
  \bibinfo {author} {\bibfnamefont {Thomas}\ \bibnamefont {Helfer}}, \bibinfo
  {author} {\bibfnamefont {Emanuele}\ \bibnamefont {Berti}}, \bibinfo {author}
  {\bibfnamefont {Pedro~G.}\ \bibnamefont {Ferreira}}, \ and\ \bibinfo {author}
  {\bibfnamefont {Lam}\ \bibnamefont {Hui}},\ }\bibfield  {title} {\enquote
  {\bibinfo {title} {{Relativistic drag forces on black holes from scalar dark
  matter clouds of all sizes}},}\ }\href {\doibase
  10.1103/PhysRevD.108.L121502} {\bibfield  {journal} {\bibinfo  {journal}
  {Phys. Rev. D}\ }\textbf {\bibinfo {volume} {108}},\ \bibinfo {pages}
  {L121502} (\bibinfo {year} {2023})},\ \Eprint
  {http://arxiv.org/abs/2305.10492} {arXiv:2305.10492 [gr-qc]} \BibitemShut
  {NoStop}%
\bibitem [{\citenamefont {Aurrekoetxea}\ \emph
  {et~al.}(2024{\natexlab{a}})\citenamefont {Aurrekoetxea}, \citenamefont
  {Clough}, \citenamefont {Bamber},\ and\ \citenamefont
  {Ferreira}}]{Aurrekoetxea:2023jwk}%
  \BibitemOpen
  \bibfield  {author} {\bibinfo {author} {\bibfnamefont {Josu~C.}\ \bibnamefont
  {Aurrekoetxea}}, \bibinfo {author} {\bibfnamefont {Katy}\ \bibnamefont
  {Clough}}, \bibinfo {author} {\bibfnamefont {Jamie}\ \bibnamefont {Bamber}},
  \ and\ \bibinfo {author} {\bibfnamefont {Pedro~G.}\ \bibnamefont
  {Ferreira}},\ }\bibfield  {title} {\enquote {\bibinfo {title} {{Effect of
  Wave Dark Matter on Equal Mass Black Hole Mergers}},}\ }\href {\doibase
  10.1103/PhysRevLett.132.211401} {\bibfield  {journal} {\bibinfo  {journal}
  {Phys. Rev. Lett.}\ }\textbf {\bibinfo {volume} {132}},\ \bibinfo {pages}
  {211401} (\bibinfo {year} {2024}{\natexlab{a}})},\ \Eprint
  {http://arxiv.org/abs/2311.18156} {arXiv:2311.18156 [gr-qc]} \BibitemShut
  {NoStop}%
\bibitem [{\citenamefont {Aurrekoetxea}\ \emph
  {et~al.}(2024{\natexlab{b}})\citenamefont {Aurrekoetxea}, \citenamefont
  {Marsden}, \citenamefont {Clough},\ and\ \citenamefont
  {Ferreira}}]{Aurrekoetxea:2024cqd}%
  \BibitemOpen
  \bibfield  {author} {\bibinfo {author} {\bibfnamefont {Josu~C.}\ \bibnamefont
  {Aurrekoetxea}}, \bibinfo {author} {\bibfnamefont {James}\ \bibnamefont
  {Marsden}}, \bibinfo {author} {\bibfnamefont {Katy}\ \bibnamefont {Clough}},
  \ and\ \bibinfo {author} {\bibfnamefont {Pedro~G.}\ \bibnamefont
  {Ferreira}},\ }\bibfield  {title} {\enquote {\bibinfo {title}
  {{Self-interacting scalar dark matter around binary black holes}},}\ }\href
  {\doibase 10.1103/PhysRevD.110.083011} {\bibfield  {journal} {\bibinfo
  {journal} {Phys. Rev. D}\ }\textbf {\bibinfo {volume} {110}},\ \bibinfo
  {pages} {083011} (\bibinfo {year} {2024}{\natexlab{b}})},\ \Eprint
  {http://arxiv.org/abs/2409.01937} {arXiv:2409.01937 [gr-qc]} \BibitemShut
  {NoStop}%
\bibitem [{\citenamefont {Brizuela}\ \emph {et~al.}(2024)\citenamefont
  {Brizuela}, \citenamefont {de~Cesare},\ and\ \citenamefont
  {Oficial}}]{Brizuela:2024smr}%
  \BibitemOpen
  \bibfield  {author} {\bibinfo {author} {\bibfnamefont {David}\ \bibnamefont
  {Brizuela}}, \bibinfo {author} {\bibfnamefont {Marco}\ \bibnamefont
  {de~Cesare}}, \ and\ \bibinfo {author} {\bibfnamefont {Araceli~Soler}\
  \bibnamefont {Oficial}},\ }\bibfield  {title} {\enquote {\bibinfo {title}
  {{Perturbations of bimetric gravity on most general spherically symmetric
  spacetimes}},}\ }\href {\doibase 10.1103/PhysRevD.109.124060} {\bibfield
  {journal} {\bibinfo  {journal} {Phys. Rev. D}\ }\textbf {\bibinfo {volume}
  {109}},\ \bibinfo {pages} {124060} (\bibinfo {year} {2024})},\ \Eprint
  {http://arxiv.org/abs/2402.15327} {arXiv:2402.15327 [gr-qc]} \BibitemShut
  {NoStop}%
\bibitem [{\citenamefont {Hassan}\ and\ \citenamefont
  {Rosen}(2012)}]{Hassan:2011zd}%
  \BibitemOpen
  \bibfield  {author} {\bibinfo {author} {\bibfnamefont {S.~F.}\ \bibnamefont
  {Hassan}}\ and\ \bibinfo {author} {\bibfnamefont {Rachel~A.}\ \bibnamefont
  {Rosen}},\ }\bibfield  {title} {\enquote {\bibinfo {title} {{Bimetric Gravity
  from Ghost-free Massive Gravity}},}\ }\href {\doibase
  10.1007/JHEP02(2012)126} {\bibfield  {journal} {\bibinfo  {journal} {JHEP}\
  }\textbf {\bibinfo {volume} {02}},\ \bibinfo {pages} {126} (\bibinfo {year}
  {2012})},\ \Eprint {http://arxiv.org/abs/1109.3515} {arXiv:1109.3515
  [hep-th]} \BibitemShut {NoStop}%
\bibitem [{\citenamefont {Camilloni}\ \emph {et~al.}(2020)\citenamefont
  {Camilloni}, \citenamefont {Grignani}, \citenamefont {Harmark}, \citenamefont
  {Oliveri},\ and\ \citenamefont {Orselli}}]{Camilloni:2020hns}%
  \BibitemOpen
  \bibfield  {author} {\bibinfo {author} {\bibfnamefont {F.}~\bibnamefont
  {Camilloni}}, \bibinfo {author} {\bibfnamefont {G.}~\bibnamefont {Grignani}},
  \bibinfo {author} {\bibfnamefont {T.}~\bibnamefont {Harmark}}, \bibinfo
  {author} {\bibfnamefont {R.}~\bibnamefont {Oliveri}}, \ and\ \bibinfo
  {author} {\bibfnamefont {M.}~\bibnamefont {Orselli}},\ }\bibfield  {title}
  {\enquote {\bibinfo {title} {{Moving away from the Near-Horizon Attractor of
  the Extreme Kerr Force-Free Magnetosphere}},}\ }\href {\doibase
  10.1088/1475-7516/2020/10/048} {\bibfield  {journal} {\bibinfo  {journal}
  {JCAP}\ }\textbf {\bibinfo {volume} {10}},\ \bibinfo {pages} {048} (\bibinfo
  {year} {2020})},\ \Eprint {http://arxiv.org/abs/2007.15665} {arXiv:2007.15665
  [gr-qc]} \BibitemShut {NoStop}%
\bibitem [{\citenamefont {Camilloni}\ \emph {et~al.}(2021)\citenamefont
  {Camilloni}, \citenamefont {Grignani}, \citenamefont {Harmark}, \citenamefont
  {Oliveri},\ and\ \citenamefont {Orselli}}]{Camilloni:2020qah}%
  \BibitemOpen
  \bibfield  {author} {\bibinfo {author} {\bibfnamefont {Filippo}\ \bibnamefont
  {Camilloni}}, \bibinfo {author} {\bibfnamefont {Gianluca}\ \bibnamefont
  {Grignani}}, \bibinfo {author} {\bibfnamefont {Troels}\ \bibnamefont
  {Harmark}}, \bibinfo {author} {\bibfnamefont {Roberto}\ \bibnamefont
  {Oliveri}}, \ and\ \bibinfo {author} {\bibfnamefont {Marta}\ \bibnamefont
  {Orselli}},\ }\bibfield  {title} {\enquote {\bibinfo {title} {{Force-free
  magnetosphere attractors for near-horizon extreme and near-extreme limits of
  Kerr black hole}},}\ }\href {\doibase 10.1088/1361-6382/abdf70} {\bibfield
  {journal} {\bibinfo  {journal} {Class. Quant. Grav.}\ }\textbf {\bibinfo
  {volume} {38}},\ \bibinfo {pages} {075022} (\bibinfo {year} {2021})},\
  \Eprint {http://arxiv.org/abs/2007.15662} {arXiv:2007.15662 [gr-qc]}
  \BibitemShut {NoStop}%
\bibitem [{\citenamefont {Mart{\'\i}n-Garc{'i}a}()}]{xact}%
  \BibitemOpen
  \bibfield  {author} {\bibinfo {author} {\bibfnamefont {J.~M.}\ \bibnamefont
  {Mart{\'\i}n-Garc{'i}a}},\ }\href {http://www.xact.es/} {\enquote {\bibinfo
  {title} {xact: Efficient tensor computer algebra for the wolfram languag},}\
  }\BibitemShut {NoStop}%
\bibitem [{\citenamefont {Brizuela}\ \emph {et~al.}(2009)\citenamefont
  {Brizuela}, \citenamefont {Martin-Garcia},\ and\ \citenamefont
  {Mena~Marugan}}]{Brizuela:2008ra}%
  \BibitemOpen
  \bibfield  {author} {\bibinfo {author} {\bibfnamefont {David}\ \bibnamefont
  {Brizuela}}, \bibinfo {author} {\bibfnamefont {Jose~M.}\ \bibnamefont
  {Martin-Garcia}}, \ and\ \bibinfo {author} {\bibfnamefont {Guillermo~A.}\
  \bibnamefont {Mena~Marugan}},\ }\bibfield  {title} {\enquote {\bibinfo
  {title} {{xPert: Computer algebra for metric perturbation theory}},}\ }\href
  {\doibase 10.1007/s10714-009-0773-2} {\bibfield  {journal} {\bibinfo
  {journal} {Gen. Rel. Grav.}\ }\textbf {\bibinfo {volume} {41}},\ \bibinfo
  {pages} {2415--2431} (\bibinfo {year} {2009})},\ \Eprint
  {http://arxiv.org/abs/0807.0824} {arXiv:0807.0824 [gr-qc]} \BibitemShut
  {NoStop}%
\bibitem [{\citenamefont {Bonanos}()}]{rgtc}%
  \BibitemOpen
  \bibfield  {author} {\bibinfo {author} {\bibfnamefont {S.}~\bibnamefont
  {Bonanos}},\ }\href {http://www.inp.demokritos.gr/~sbonano/RGTC/} {\enquote
  {\bibinfo {title} {Riemannian geometry \& tensor calculus (rgtc) in
  mathematica},}\ }\BibitemShut {NoStop}%
\end{thebibliography}%

\end{document}